\documentclass[printer]{aa}
\usepackage{graphics,epsfig,lscape,txfonts}
\usepackage{natbib,aas_macros}
\bibpunct{(}{)}{;}{a}{}{,}

\def\cm-2{cm$^{-2}$}

\def\msun{M$_{\odot}$}

\def\chandra{{\it Chandra}}

\def\xmm{{XMM-Newton}}
\def\m31{\object{M~31}}
\def\me33{\object{M~33}}

\newcommand{\ergs}[1]{$\times 10^{#1}$ \hbox{erg s$^{-1}$}}

\newcommand{\hcm}[1]{$\times 10^{#1}$ cm$^{-2}$}
\newcommand{\ohcm}[1]{$10^{#1}$ cm$^{-2}$}

\newcommand{\nh}{\hbox{$N_{\rm H}$}}

\begin{document}

   \title{Optical novae: the major class of supersoft X-ray sources in M~31\thanks{Partly 
   based on observations with XMM-Newton, an ESA Science Mission 
    with instruments and contributions directly funded by ESA Member
    States and NASA, and on observations obtained with the Wendelstein
    Observatory of the Universit\"atssternwarte M\"unchen}
}

   \author{W.~Pietsch\inst{1} \and 
           J.~Fliri\inst{2}\thanks{Visiting astronomer at the German-Spanish
	   Astronomical Center, Calar Alto, operated by the Max-Planck-Institut
	   f\"ur Astronomie, Heidelberg, jointly with the Spanish National
	   Commission for Astronomy} \and
	   M.~J.~Freyberg\inst{1} \and
	   J.~Greiner\inst{1} \and
	   F.~Haberl\inst{1} \and
	   A.~Riffeser\inst{1,2 \star\star} \and
	   G.~Sala\inst{1,3}
          }
\institute{Max-Planck-Institut f\"ur extraterrestrische Physik, Giessenbachstra\ss e, 
           85741 Garching, Germany 
           \and 
	   Universit\"atssternwarte M\"unchen, Scheinerstra\ss e, 81679
	   M\"unchen, Germany 
	   \and
	   Institut d'Estudis Espacials de Catalunya (ICE-CSIC), Campus UAB, Facultat de
           Ciencies, 08193 Bellaterra, Spain
           }
     
     \offprints{W.~Pietsch, \email{wnp@mpe.mpg.de}}

   \date{Received / Accepted }

	\abstract{We searched for X-ray counterparts of optical novae detected in \m31\ and \me33.  
         We combined an optical nova catalogue from the WeCAPP survey with optical novae
         reported in the literature and correlated them with the most recent X-ray catalogues
         from ROSAT, \xmm\ and \chandra, and -- in addition -- searched for nova correlations 
         in archival data. We report 21 X-ray counterparts for novae in \m31 -- mostly identified as 
	 supersoft sources (SSS) by their hardness ratios -- and
         two in \me33. Our sample more than triples the number of known optical novae with supersoft 
	 X-ray phase. 
         Most of the counterparts are covered in several observations allowing 
         us to constrain their X-ray light curves. Selected brighter sources were
         classified by their \xmm\ EPIC spectra. We use the well determined 
	 start time of the SSS state in two novae
	 to estimate the hydrogen mass ejected in the outburst to $\sim10^{-5}M_{\odot}$ and 
	 $\sim10^{-6}M_{\odot}$, respectively.
         The supersoft X-ray phase of at least 15\% of the novae starts within a year. 
	 At least one of the novae shows a SSS state lasting 6.1 years after
	 the optical outburst. Six of the SSSs turned on between 3 and 9 years 
	 after the optical discovery of the outburst and may be interpreted as 
	 recurrent novae. If
	 confirmed, the detection of a delayed SSS phase turn-on may be used as a new method 
	 to classify novae as recurrent. At the moment, the new method yields a
	 ratio of recurrent novae to classical novae of 0.3 which is in agreement (within the
	 errors) with previous works.
	
} 
\maketitle

\section{Introduction}
After the first discovery of X-ray emission from nova GQ Mus 
463 days after outburst \citep{1984ApJ...287L..31O} 
and its subsequent detection by ROSAT nine years after outburst
\citep{1993Natur.361..331O}
it was widely believed that post-novae
should go through an extended phase of residual hydrogen burning
and appear as supersoft X-ray source (SSS). Combining the burning rate
of 10$^{-7}$ \msun/yr with the assumption that there should be
of order 10$^{-4}$ \msun\ in the shell to ignite hydrogen burning,
a duration of the burning phase of 10$^3$ to 10$^4$ yrs had
been predicted 
\citep{1985ApJ...294..263M,sta89}.
However, supersoft X-ray emission has been detected so far for
only five other novae: V1974 Cyg detected as SSS for about 400 days starting
from 250 days after outburst \citep{1996ApJ...456..788K};
Nova LMC 1995 from 200 to more than 2100 days after outburst 
\citep{1999A&A...344L..13O,2003ApJ...594..435O}, but had almost faded after 
8 years \citep{2004RMxAC..20..182O}; 
V382 Vel, with 
an intense SSS component detected about 180 and 222 days after outburst, 
but disappearing by day 268 
\citep{2002MNRAS.333L..11O,2002AIPC..637..377B};
V1494 Aql about 250 days after outburst \citep{2003ApJ...584..448D} and no
later X-ray observations reported;
V4743 Sgr about 180 days after outburst, and still appearing as SSS at day 740
after outburst \citep{2003ApJ...594L.127N,2004IAUC.8435....2O}.
In a systematic search through all ROSAT pointed
observations covering 39 novae over a time span up to 10 yrs,
no further supersoft phase was discovered 
\citep{2001A&A...373..542O}.
Although only X-ray observations can provide direct insight into the hot
post-outburst white dwarf, ultraviolet emission lines arising from the
ionization of the ejecta by the central X-ray source reflect the presence
of on-going hydrogen burning on the white dwarf surface. Several works
have used this indirect indicator to determine the turn-off of classical
novae from IUE observations 
\citep{1996ApJ...463L..21S,1998A&AS..129...23G,2001AJ....121.1126V},
showing in all cases turn-off times
shorter than expected.
The short duration of the H-burning phase derived from observations could
be explained by a small post-outburst envelope mass, suggesting the
presence of some extra mass loss mechanism acting after the nova outburst,
i.e., a thick wind \citep{1994ApJ...437..802K}
or a common envelope \citep{1985ApJ...294..263M}. 
In fact, post-nova white dwarf envelopes with steady
H-burning are stable only for masses smaller than about $10^{-5}~\rm{M}_{\odot}$
\citep{1998ApJ...503..381T,SH2005a},
which also suggests that
instabilities in envelopes with larger masses could contribute to get rid
of the mass excess.
With the small
Galactic novae rate there was no substantial improvement
in our understanding of these processes during the last 5 yrs
with \chandra\ and \xmm\ observations.

However, observing the nova population in \m31\ has the advantage that
this as well as several
additional questions can be attacked much more easily as compared to the local
sources (including those in the Magellanic Clouds): (i) What is
the spatial distribution
over the galaxy and are there possible correlations with different environment?
(ii) What is the size of the population including the fraction of sources
detected during their supersoft X-ray emission phase?
(iii) What is the variability pattern? (iv) Are there correlations
between the optical and X-ray properties?

There have been many surveys for optical novae in \m31\ starting with the early
work of \citet{1929ApJ....69..103H}, who used the novae to establish the
distance of \m31\ and already estimated a yearly nova rate of 30 and found that
novae are most frequent in the nuclear area. Novae were detected by comparing
plates taken at different times. 
However, many were missed due to the sparse sampling and the
shortness of the nova outbursts. When it was noticed that novae stayed bright in
H$\alpha$\ for a longer time this band was used for \m31\ nova searches
\citep[see e.g.][]{1990ApJ...356..472C}. With this method many nova candidates
were detected with only rough knowledge of the date of outburst as well as 
duration. With the start
of the pixellensing surveys of the center area of \m31\ many novae were detected
as a by-product with good sampling of the outbursts which led to well defined
outburst dates and decay time scales of many novae simultaneously. In Sect. 2 we
report nova detections from one of these programs. For the search of X-rays from
novae in \m31\ we combine this nova list with novae reported in the literature. 
This nova list contains about 10--20 novae per year prior to the \xmm\ and
\chandra\ observations and may be 30\% to 60\% complete, while in the years 
before 1990 (novae to shine up during the ROSAT observations) typically less 
than 5 novae were reported.

\m31\ \citep[distance 780 kpc,][]{1998AJ....115.1916H,1998ApJ...503L.131S}
with its moderate Galactic foreground absorption  \citep[\nh = 6.66\hcm{20},
][]{1992ApJS...79...77S} is an ideal target to search for X-ray emission from 
optical novae. 
ROSAT has observed the full disk of the \m31\ galaxy (about 6.5 deg$^2$) twice.
A ROSAT PSPC mosaic of 6 contiguous pointings with an exposure time of
25 ks each was performed in July 1991 
\citep[first \m31\ survey;][hereafter SHP97]{1997A&A...317..328S}.
A second survey
was made in July/August 1992 and January and July 1993  
\citep[][hereafter SHL2001]{2001A&A...373...63S}.
Only one recent nova (which erupted in 1990) in \m31\ was reported
to coincide with a
cataloged ROSAT source \citep{2002A&A...389..439N}.
The population of SSS in \m31\ has been studied by
\citet{1996LNP96.472...75G,2004ApJ...610..261G}, in particular their variability.
One of the surprising results was that more fading than
rising sources have been found. Coincidentally, one of
these faders was the above mentioned nova 
\citep[RX~J0044.0+4118;][]{2002A&A...389..439N}. 
This led to the  speculation that the
difference in the numbers of faders and risers is due to a fraction
of classical novae for which the X-ray rising phase could be
much shorter than the fading phase.
Based on the until then known durations of the supersoft
X-ray phases this explanation was considered unlikely.
Also, the global (bulge+disk) nova rate of
$\sim$37 nova per year in \m31\ \citep{2001ApJ...563..749S}
combined with the short duration of the ROSAT survey
did not suggest more than two novae among the two dozen ROSAT
SSSs in \m31\ when taking into account
the wide spread locations of the SSSs over the \m31\ disk.
Similarly, recurrent nova were not expected to contribute
to the observed SSS sample since
the outburst rate of recurrent novae in \m31\ has been estimated to be only
10\% of the rate of classical novae \citep{1996ApJ...473..240D}.

The \xmm\ survey of \m31\ has identified 856 X-ray sources \citep[][hereafter PFH2005]{PFH2005}
analyzing all observations  in the \xmm\ archive which
overlap at least in part with the optical $D_{25}$ extent of the galaxy.
Among them are 18 SSSs defined by $HR1 < 0$ and $HR2 - EHR2 < -0.4$. 
Based on count rates in energy bands 1 to 3 (0.2--0.5 keV, 0.5--1.0 keV, 1.0--2.0 keV), 
$HRi$ and $EHRi$ are defined as 

$HRi = \frac{B_{i+1} - B_{i}}{B_{i+1} + B_{i}}$  and
$EHRi = 2  \frac{\sqrt{(B_{i+1} EB_{i})^2 + (B_{i} EB_{i+1})^2}}{(B_{i+1} + B_{i})^2}$, 

\noindent for {\it i} = 1,2, 
where $B_{i}$ and $EB_{i}$ denote count rates and corresponding errors in
band {\it i}.

In addition there are X-ray source catalogues from two deep observations 
of the \m31\ center area with \chandra\ ACIS~S
\citep[][hereafter DKG2004]{2004ApJ...610..247D} 
and HRC~I \citep[][hereafter K2002]{2002ApJ...578..114K} sensitive for
SSS emission. 
Several \chandra\ ACIS~I observations 
\citep[see e.g.][]{2002ApJ...577..738K} are not
sensitive for the detection of SSS. However,  some short
observations (\chandra\ ACIS~S and HRC~I) in the archive can also be
searched for optical novae. 

In X-rays novae may be visible as bright SSS for several years; 
therefore many missed novae may show up as SSS. Also recurrent novae, 
optically known as novae from previous or later outbursts 
may show up in X-rays. 
Out of this reason and due to the uncertain outburst dates as discussed above, 
derived times since outburst have to be taken with care.
For all \m31\ novae the distance is about the same. However, extinction within \m31\
may hamper the interpretation of the supersoft emission. 

With  the much larger emphasis on the bulge of \m31\ with its high
concentration of sources it was interesting to reconsider
the detection rate of novae which will be presented in Sect. 3 together with
a discussion of the individual objects in subsections.
Finally, we discuss the results and demonstrate the
wealth of information that can be expected from a continuing optical and X-ray
nova survey in the center area of \m31.
 
\section{\m31\ optical nova catalogue}\label{optnova}
\begin{figure}
   \resizebox{\hsize}{!}{\includegraphics[]{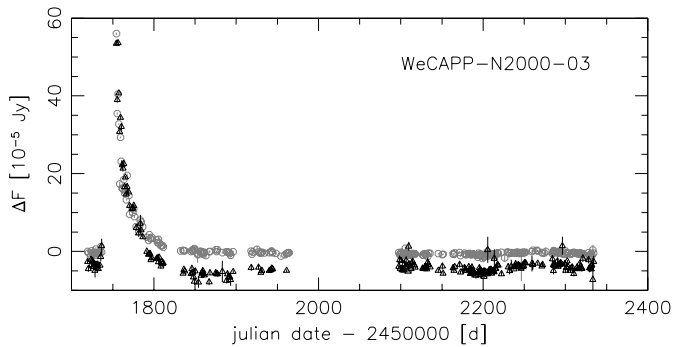}}
     \caption[]{
     Optical light curve of WeCAPP-N2000-03 in R (grey circles) and I 
     (black triangles). The I band data show a negative bias as the reference image
     contained nova epochs.  
     }
    \label{nova_opt} 
\end{figure}
The optical novae used for cross-correlation with the X-ray data,
result in part from two years of observations (2000, June 23
to 2002, February 28)  
of the central part of \m31\ by the still ongoing Wendelstein Calar Alto Pixellensing
Project (WeCAPP, \citealt{2001A&A...379..362R}). 
WeCAPP monitors a $17.2\arcmin \times 17.2\arcmin$ field
centered on the nucleus with the 0.8~m telescope at Wendelstein
Observatory (Germany) and the 1.23~m telescope at Calar Alto 
Observatory (Spain) continuously since 1997. The observations are 
carried out in $R$ and $I$ filters close to the Kron-Cousins system. 
The data are collected with a rather dense time coverage 
(up to 335 and 310 epochs in the $R$- and $I$-bands, respectively).
Data were reduced using the WeCAPP reduction pipeline {\sf mupipe}, 
which implements an image subtraction technique 
\citep{1998ApJ...503..325A} to overcome the crowding effects and allow proper
photometry of variable sources in the central bulge of \m31\ \citep{2003ApJ...599L..17R}.
The pipeline combines the standard CCD reduction (including de-biasing,
flat-fielding and filtering of cosmic ray events), position alignment, 
photometric calibration, and restoration of damaged pixels
with full error propagation for each pixel of the CCD frame \citep{2002A&A...381.1095G}.
After the point spread-functions (PSF) of a high $S/N$ reference frame and 
the stacked frames are matched, the reference image is subtracted from
all other frames, generating difference images for each observation night.
PSF photometry of each pixel in the difference frames finally results
in 4 $\times 10^6$ pixel light curves with appropriate error bars,
each of them represents the temporal variability of the flux inside
the PSF centered on the particular pixel.
In the full WeCAPP data set, 23770 variable sources were detected,
most of them being Long Period Variables \citep{wecapp04}.
The 1$\sigma$ error radius of the astrometric solution is 0\farcs16. 
Novae are amongst the brightest variable sources in the data set. They
therefore could be detected by a simple but effective algorithm. 
As first cut two consecutive data points in the light curve were 
required that exceed a difference flux level $\Delta F_R$ of 
$6\times10^{-5} \rm{Jy}$ above the baseline
(corresponding to a detection limit of
  $M_R = -2.5 \log \left(\frac{6\times10^{-5}~{\rm Jy}}{F_{{\rm Vega},R}}\right) \approx 19.3~{\rm mag}$
  with $F_{{\rm Vega},R}=3060~{\rm Jy}$ being the flux of Vega in the 
  $R$-band). 
The light curves fulfilling this criterion are than
inspected visually to extract the nova candidates. A catalogue of all 40 novae,
detected in the survey will be published separately. The outburst of about half
the sample occured after the X-ray observations. As an example for a nova 
which correlates with a time variable SSS detected by \xmm\ and \chandra,
we show the optical light curve of WeCAPP-N2000-03 (Fig.~\ref{nova_opt}).

We combined the WeCAPP nova list with novae from other microlensing 
surveys of \m31: the AGAPE survey \citep{2004A&A...421..509A}, 
the POINT-AGAPE PACN survey \citep{2004MNRAS.351.1071A,2004MNRAS.353..571D},  
the Nainital Microlensing Survey \citep{2004A&A...415..471J} and
the survey by \citet[][hereafter TC96]{1996AJ....112.2872T}. 
We added novae from IAU circulars 
and astronomical telegrams (ATEL). We included 
novae from the H$\alpha$ searches of \citet[][hereafter SI2001]{2001ApJ...563..749S},
\citet[][hereafter RJC99, nova and nova candidate lists\footnote{available at
http://www.noao.edu/outreach/rsbe/nova.html}]{1999AAS...195.3608R},
\citet{1990ApJ...356..472C} and \citet[][hereafter CFN87]{1987ApJ...318..520C}. 
We added the lists by Sharov and colleagues \citep{1991Ap&SS.180..273S,1992Ap&SS.188..143S,1992Ap&SS.190..119S,1993AstL...19..230S,1994AstL...20...18S,
1994AstL...20..711S,1995AstL...21..579S,1996AstL...22..680S,1997AstL...23..540S,1998AstL...24..445S,1998AstL...24..641S,2000AstL...26..433S}. 
For two
recurrent novae we use the naming convention provided in the General Catalogue of Variable Stars \citep{2004yCat.2250....0S}.
We refrained from using positions of earlier nova catalogues for the cross-correlation as 
nova positions in the earlier catalogues are only determined to 0.1 arcmin or worse. This might 
lead to many spurious correlations specifically in the central region of \m31, which is crowded 
with novae and X-ray sources. 

\section{\m31\ and \me33\ optical novae detected with \xmm, Chandra and ROSAT}
We compared the \m31\ optical nova catalogue with archival and/or published 
\xmm, \chandra\ and ROSAT data.  

For \xmm\ 
\citep{2001A&A...365L...1J}  we reanalyzed the same archival EPIC 
\citep{2001A&A...365L..18S,2001A&A...365L..27T} observations that were used by
PFH2005
for the creation of the \m31\ source catalogue, i.e. pointings c1 to c4 (June 2000 
to January 2002) to the
galaxy center, n1 to n3 (January 2002 and June 2002) to the northern disk, s1 and s2 
(January 2002) to the southern disk and h4 (January 2002) to the northwest halo 
(see Table 1 of PFH2005 for details). The observations
were performed in the full frame mode using medium or thin filter with low background
exposure times of about 10 to 50 ks. We correlated the optical nova catalogue with  
the sources from the PFH2005 catalogue and determined luminosities or upper limits
for nova candidates for each observation. We give luminosities for the sources that
are detected with at least 2$\sigma$ significance in the (0.2--1.0) keV band combining
all EPIC instruments. Upper limits are 3$\sigma$ determined from the more sensitive
EPIC pn camera when possible. For bright sources we analyzed the X-ray spectra.

For \chandra\ we correlated the optical nova catalogue with sources in
the \m31\ center area presented by DKG2004 and K2002 based on 
a 37.7 ks ACIS~S (ObsID 1575) and a 46.8 ks HRC~I observation (ObsID 1912), 
respectively. These
observations were performed in October 2001, between the third (c3) and forth (c4)
\xmm\ observation of the \m31\ center. For novae not reported in these \chandra\ 
catalogues but detected by \xmm\ we determined luminosities (if detected with a 
significance greater 2$\sigma$, else we calculated 
3$\sigma$ upper limits). In addition, we
searched in further short archival HRC~I and ACIS~S observations for nova detections
and report nova correlations if more than 4 counts are detected. 
In only one of the many $\sim$1.2~ks HRC~I observations a new nova candidate brighter than 
this limit was detected (Nova WeCAPP-N2002-01, see below). Unfortunately, also
upper limits determined from these short observations do not constrain nova
light curves and we therefore mostly restrain from reporting these limits.  
A search in the ACIS~I 
catalogue by \citet{2002ApJ...577..738K} and the HRC~I snap shot catalogue by
\citet{williams2004b} yielded no additional nova candidates.

For ROSAT we correlated the optical nova catalogue with the source catalogues from 
the first and second PSPC \m31\ survey (SHP97 and SHL2001) and with the HRI catalogue
of \citet[][hereafter PFJ93]{1993ApJ...410..615P}. In addition we searched in archival ROSAT PSPC and 
HRI observations for further nova correlations. From HRI information alone we can not 
decide if the proposed counterpart is a SSS.

\begin{figure}
   \resizebox{\hsize}{!}{\includegraphics[]{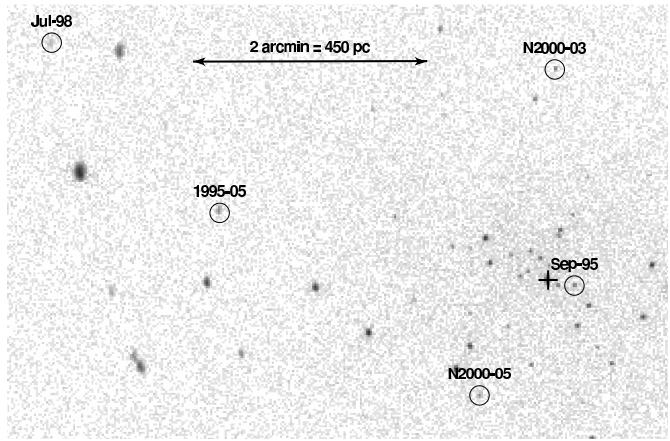}}
     \caption[]{Part of the
     \chandra\ HRC~I image of observation 1912 used for the source catalogue of
     K2002. Circles with 5" radius indicate nova positions. The cross indicates
     the \m31\ center, the aim point of the observation.
     }
    \label{Kaaret} 
\end{figure}

In the \xmm\ and \chandra\ data, correlations with 17 optical novae in \m31\ were detected.  
We rejected three correlations where the X-ray counterpart was classified as hard by PFH2005:
the sources [PFH2005]~299, 412 and 601 which correlate with [SI2001]~1997-06, [CFN87]~26 and Nova~21
\citep{1998AstL...24..445S}, respectively. While the first two may be chance coincidences in the 
densely populated central area of \m31, Nova~21 -- which also correlates with the hard ROSAT source 
[SHL2001]~306 -- has been classified unique from its light curve by \citet{1998AstL...24..445S}
and may not be a nova at all.
Five optical novae correlate with the \chandra\ ACIS~S catalogue of
DKG2004, six with the  \chandra\ HRC~I catalogue of K2002. Figure~\ref{Kaaret} 
demonstrates the detection of five novae near the \m31\ center in HRC~I 
observation 1912.
Nova WeCAPP-N2002-01 is detected with 6 counts in the 1 ks archival HRC~I observation
2906 and [SI2001] 1997-06 11 counts in the 5.2~ks HRC~I observation 268. Eight
novae are contained in the \xmm\ source catalogue of PFH2005 and are detected in
at least one of the contributing observations. 
All but one (the probable symbiotic [PFH2005] 395) \xmm\ and \chandra\ HRC~I nova candidates have been 
classified as SSS  by 
DKG2004 and PFH2005. There are only three novae ([SI2001] 1997-06, WeCAPP-N2001-08 
and WeCAPP-N2002-01)
that are just identified by positional coincidence in \chandra\ HRC~I observations
and not also by their supersoft spectrum. 

\citet{2004MNRAS.351.1071A} give correlation results of the POINT-AGAPE survey list
of variable stars in \m31\ with 13 known novae. For four of them they propose X-ray
counterparts. Novae EQ J004244+411757 and CXOM31~J004318.5+410950 are confirmed as
X-ray emitting novae (novae WeCAPP-N2000-03 and WeCAPP-N2001-08).
WeCAPP-N2001-12 is detected in X-rays as described below. However, the hard X-ray 
transient [OBT2001]~3 \citep{2001A&A...378..800O} is not the counterpart. 
The forth X-ray candidate
(CXOM31~J004222.3+411333) is more than 10\arcsec\ from the position of 
EQ~J004242+411323 \citep[given in][]{1999IAUC.7236....1J} and about
8\arcsec\ from the position given in  \citet{2001A&A...378..800O} and
most probably only a chance coincidence in this crowded source region.
The X-ray source is reported in the \xmm\ catalogue as [PFH2005]~255
as a hard source that was also detected in the EINSTEIN and ROSAT HRI and 
PSPC surveys.

In the ROSAT PSPC and HRI data, correlations with five respectively two optical 
novae in \m31\ were detected.  
Three sources
correlate with sources of the first \m31\ ROSAT catalogue (SHP97) and another two with 
sources from the second (SHL2001). The ROSAT PSPC sources have a component in the soft
band indicative for SSS. 
The second ROSAT survey contains observations collected 
in three epochs. For the ROSAT nova candidates we determined luminosities for each of the 
epochs (SII-E1 to E3) and assume average Julian Dates (JD) of 2448840.5, 2448990.5 and 
2449190.5, respectively. For this purpose we only merged observations of the different
epochs where the source position was less than 15\arcmin\ off-axis.
For ROSAT sources from the first survey that were not detected 
in the second survey and vice versa we derived upper limits when possible.
The ROSAT detection and upper limits of the \m31\ nova of \citet{2002A&A...389..439N}
are also indicated. One nova correlates with a source from PFJ93, another one with 
two HRI observations in July 1994. Both sources are not detected in ROSAT PSPC
and are just  identified by positional coincidence in ROSAT HRI observations
and not also by their supersoft spectrum.

Inspired by the many nova correlations in \m31\ we also checked the
\xmm\ and ROSAT catalogues from \citet[][hereafter PMH2004]{2004A&A...426...11P} and 
\citet[][hereafter HP2001]{2001A&A...373..438H} as well as the archival \chandra\ data 
of the Local Group Sc spiral \me33\ \citep[distance 795~kpc,][]{1991PASP..103..609V}
for possible nova correlations. \me33\ hosts 
less SSS and the number of known optical novae is much less. We did not find a 
correlation with an optical nova within the 408 \xmm\ sources of PMH2004. 
However, the ROSAT HRI source [HP2001]~93 clearly correlates 
with a nova. An additional \me33\ nova correlation with a SSS was detected in the
\chandra\ ACIS S observation 786.

As mentioned above, we detected three optical novae which correlate with hard sources in the 
PFH2005 catalogue. We consider two of them as chance coincidences. The number of hard sources 
in the catalog exceeds 
the number of SSS sources by a factor of 30. Therefore the number of chance coincidences
for SSS should be smaller by the same factor. This, together with the detection of the 
expected X-ray spectrum and -- in some cases -- the expected time variability, confirms the
identification of the X-ray source as optical nova counterpart. The number of sources detected 
only by the ROSAT HRI or \chandra\ HRC is much smaller than that in the PFH2005 catalogue,
position errors are similar or smaller, and therefore the number of chance coincidences is even less.
In addition, the time variability argument also holds for most of these sources, making all of 
them convincing optical nova counterparts.
 
\begin{table}
\begin{center}
\caption[]{Count rate conversion factors to unabsorbed fluxes (ECF) into the 0.2--1 keV band 
           for black body models with temperatures of 40~eV and 50~eV 
	   for different instruments and filters, including
	   a Galactic foreground absorption of 6.66\hcm{20}.}
\begin{tabular}{lrrr}
\hline\noalign{\smallskip}
\hline\noalign{\smallskip}
\multicolumn{1}{l}{Detector} & \multicolumn{1}{c}{Filter} &\multicolumn{1}{c}{40 eV} &\multicolumn{1}{c}{50 eV} \\ 
\noalign{\smallskip}
& & \multicolumn{2}{c}{($10^{-11}$ erg cm$^{-2}$ ct$^{-1}$)} \\
\noalign{\smallskip}\hline\noalign{\smallskip}
EPIC PN   & thin   & 1.77 & 1.04 \\
          & medium & 2.15 & 1.22 \\
EPIC MOS1 & thin   & 9.66 & 5.14 \\
          & medium & 11.4 & 5.94 \\
EPIC MOS2 & thin   & 9.23 & 5.00 \\
          & medium & 11.1 & 5.78 \\
Chandra ACIS~S &   & 7.25 & 5.10 \\
Chandra HRC~I  &   & 9.17 & 6.76 \\	  
ROSAT HRI &        & 29.9 & 24.9 \\
ROSAT PSPC&        & 6.42 & 6.55 \\
\noalign{\smallskip}
\hline
\noalign{\smallskip}
\end{tabular}
\label{ecf}
\end{center}
\end{table}
We calculated intrinsic luminosities or $3\sigma$ upper limits in the 0.2--1.0 keV band 
starting from the 0.2--1 keV count rates or upper limits in EPIC
and the full count rates or upper limits in the other instruments and
assuming a black body spectrum
and Galactic foreground absorption. Table~\ref{ecf} gives energy conversion factors
for the different instruments for a 40 eV and a 50 eV black body temperature. As one can see
the ECFs strongly change with the softness of the spectrum. Additional absorption within 
\m31\ would heavily change the observed count rate. An extrapolation to the bolometric 
luminosity of a nova at a time is very uncertain and this is even more so as the 
temperatures of novae may vary with time after outburst and from nova to nova and may
well correspond to a spectrum even softer than 30 eV. See also the discussion of
the luminosities derived from spectral modeling of a few individual optical nova
candidates (Sect.~\ref{1992-01}, \ref{WeCAPP-N2000-03} and \ref{Jul-98}).

In Table~\ref{novae}, the results of the 23 optical nova correlations are summarized. 
We give nova name (or month of outburst) with optical
references in column 1, optical position (J2000.0, col. 2), Julian date JD of 
``outburst
maximum" (3). We indicate, if the optical maximum is well defined (to better
than 5 days), if the maximum is most likely before or after the given epoch 
or not well defined. As X-ray information we give
name of the source (4), distance $D$ between X-ray and optical position (5),
observation number (6), JD of observation (7), days since
optical nova outburst (8), X-ray luminosity in the 0.2--1.0 keV band as
described above (9), and comments like reference for detection, nova type and 
SSS classification (10).

\begin{figure*}
   \resizebox{\hsize}{!}{\includegraphics[angle=-90,clip]{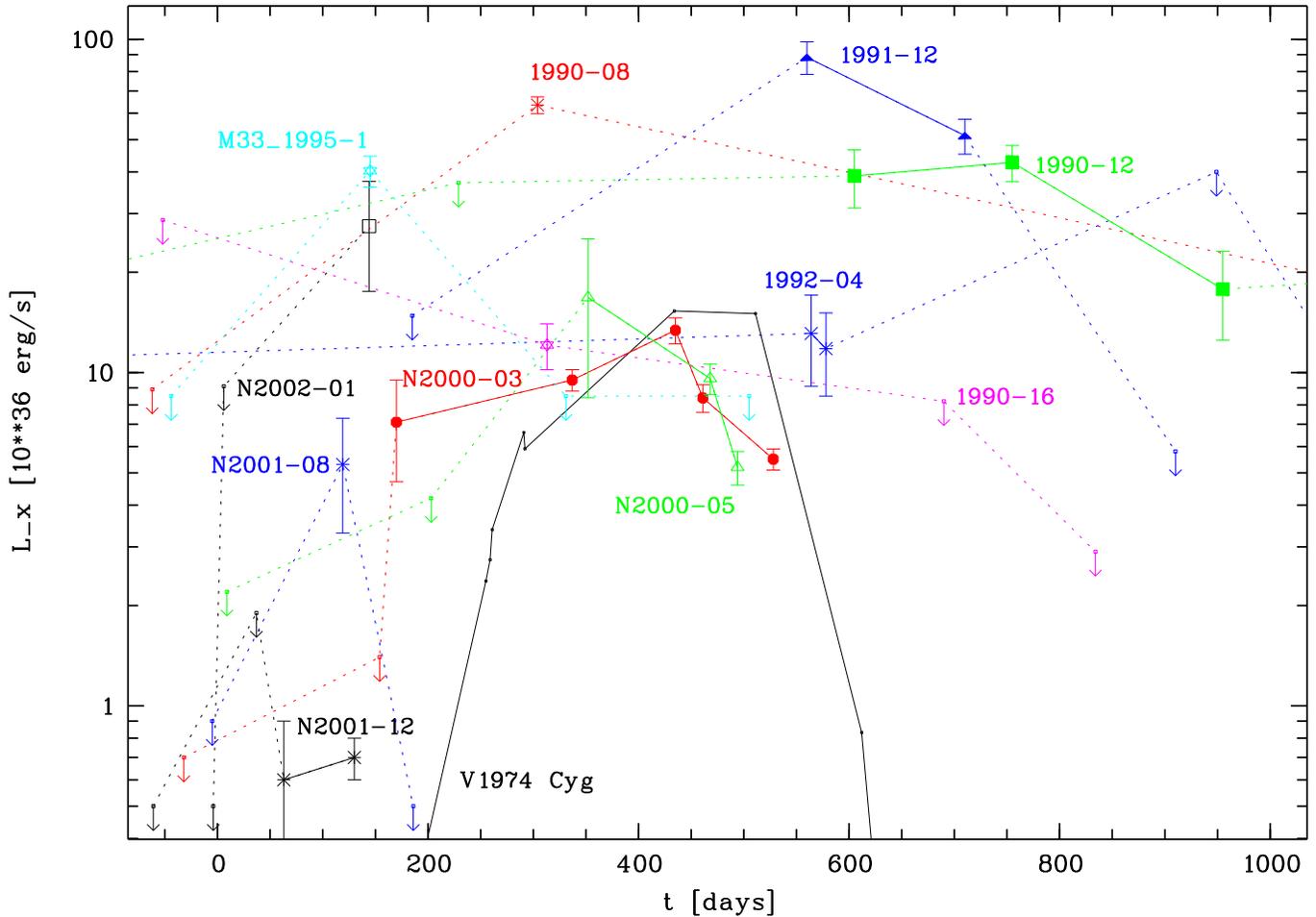}}
     \caption[]{Light curves for \m31\ and \me33\ novae that were detected within 
     1000~d after outburst. 
     Detections of individual novae are connected by solid lines, connections to upper 
     limits are marked by dashed lines.
     The light curve from nova V1974 Cyg is adapted from
     \citet{1996ApJ...456..788K} assuming
     1 cts s$^{-1}$ = 0.2 $\times 10^{36}$ erg s$^{-1}$. 
     }
    \label{m31_nova_lc} 
\end{figure*}

Some novae have been covered in several X-ray observations. Light curves are plotted in
Fig.~\ref{m31_nova_lc}. The
time variability for some of them is as expected from the ROSAT observation of
V1974 Cyg. 
However, some others seem to brighten significantly after more than 1000 days which 
we consequently interpret as recurrent novae.
The individual correlations are discussed in the following subsections starting with 
the sources from Table~\ref{novae}.

The data analysis was performed using tools in the \xmm\ Science Analysis System
(SAS) v6.1.0, 
EXSAS/MIDAS 03OCT\_EXP/03SEPpl1.2, and 
FTOOLS v5.2 software packages, the imaging application DS9 v3.0b6 together with
the funtools package,
the mission count rate simulator WebPIMMS v3.6a
and the spectral 
analysis software XSPEC v11.3.1.   

\subsection{Nova in \m31\ [SI2001]~1992-01}\label{1992-01}
\begin{figure}
   \resizebox{\hsize}{!}{\includegraphics[angle=-90,clip]{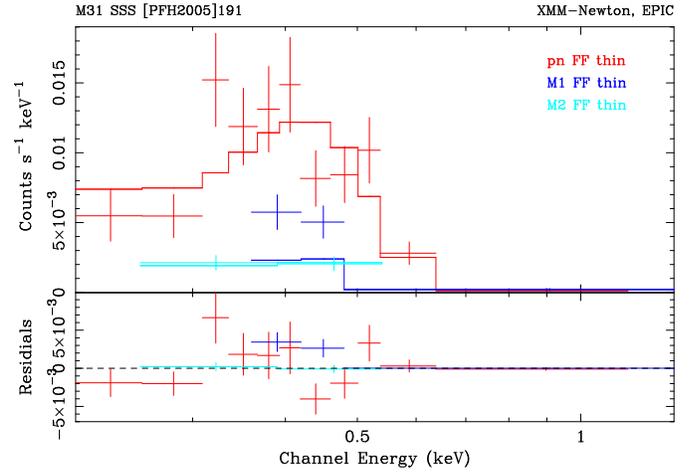}}
     \caption[]{\xmm\ EPIC spectrum of source [PFH2005]~191 
     (Nova in \m31\ [SI2001]~1992-01) for
     observation s1. The absorbed black body fit
     to the data (see Sect.~\ref{1992-01}) is shown in the upper panel. 
     }
    \label{pnspec_191} 
\end{figure}
This nova was reported by SI2001 from two H$\alpha$ images 35~d apart.
The positions are estimated by the authors to be accurate to $\sim$1\arcsec.  
The speed class is unknown.
The proposed X-ray counterpart [PFH2005]~191 is classified as SSS.  
The 3$\sigma$ X-ray positional error
is 1\farcs9, including systematics. If we take the optical position uncertainty
as 1$\sigma$, the SSS nature and the positional coincidence point at a correct
identification.

The X-ray source was first detected  $\sim$3300~d after the nova outburst 
during \xmm\ observation c4.
During this observation it was not in the field of view (FOV) of the EPIC pn
instrument. Six days later, during observation s1 to the south of 
the \m31\ center, the source was in the FOV of all EPIC cameras and about
240, 60 and 40 photons were collected by pn, MOS1 and MOS2, respectively. 
The EPIC data can best be modeled with an absorbed
black body model (\nh = ($19^{+5}_{-3}$)\hcm{20}, $kT = (48\pm11)$ eV, see
Fig.~\ref{pnspec_191}) with an unabsorbed luminosity of 6.1\ergs{37} in the
(0.2--1.0) keV band. The correction factor of $\sim$50 from absorbed to 
unabsorbed luminosity strongly depends on the assumed absorption column. 
The spectral fits show that the luminosities given in 
Table~\ref{novae} for this source are underestimated by about a factor of five. 
This is caused by the higher \nh\ determined in the spectral fit
of $\sim$20\hcm{20} while for the table only Galactic foreground absorption
(6.6\hcm{20}) was assumed.
The nova was not detected during the three \xmm\ observations
at 2743~d to 3112~d after outburst, it was not in the ACIS~S field of \chandra\ 
observation 1575. It is covered by the \chandra\ HRC~I observation 1912  
($\sim$13\arcmin\ off-axis), but not reported in the catalogue by K2002.
We determined a 3$\sigma$ upper limit indicating that the source was
not active 67~d before observation c4.

The rise of the X-ray flux of the nova rather late after the optical outburst 
could indicate that nova [SI2001]~1992-01 is a recurrent nova which had a 
new outburst after about 8 years that was not optically detected and that is 
responsible for the observed X-rays. Of course, we also can not exclude that
a physically different nova or even a classical SSS could be the counterpart.

\subsection{Nova in \m31\ WeCAPP-N2002-01}\label{WeCAPP-N2002-01}
The time of outburst maximum of the nova WeCAPP-N2002-01 \citep[also reported 
in IAU Circular by ][]{2002IAUC.7794....1F}
can only be determined to an accuracy of 8 days, as there is an observation gap
in the WeCAPP data. It is a moderately fast nova.
Its outburst occured after the \xmm\ and \chandra\ observations analyzed in
K2002, DKG2004 and PFH2005 where it was not detected. 
However, in the 1.2 ks archival \chandra\ HRC~I 
observation 2906 -- $\sim$ 144~d after the nova outburst -- we find
6 photons consistent with the \chandra\ HRMA/HRC~I point spread 
function at the position of the nova. The HRC~I provides no spectral information.
Therefore it will be important to determine from \xmm\ EPIC or \chandra\ ACIS-S
observations if the spectrum of the source is supersoft.

\subsection{Nova in \m31\ WeCAPP-N2001-12}\label{WeCAPP-N2001-12}
The time of outburst maximum of the nova WeCAPP-N2001-12 \citep[also reported 
in IAU Circular by ][]{2001IAUC.7709....3F} and in the POINT-AGAPE variable star
catalogue \citep{2004MNRAS.351.1071A}
can only be determined to an accuracy of about 10 days, as there is an observation gap
in the WeCAPP data of 5~d before and 12~d after the detection in the rising phase. 
The nova is classified as fast. 

\citet{2004MNRAS.351.1071A} propose the hard X-ray 
transient [OBT2001]~3 \citep{2001A&A...378..800O} as counterpart
which is source 287 in the PFH2005 catalogue. However, 
several points speak against this identification: (I) the 
position of this bright X-ray source is significantly offset from the nova position by 
4.4\arcsec, (II) the X-ray source was found active 430~d before the nova outburst
and was off in the \xmm\ observations 244~d and 60~d before the nova outburst, 
(III) the X-ray source showed a hard spectrum with a luminosity of 1.1\ergs{37} 
in the 0.3--10 keV band. Therefore, a neutron star or black hole X-ray transient is 
the more likely identification for [OBT2001]~3. 

During observation c4 -- 130~d after the nova outburst --  
we detect a faint SSS close to the position of [PFH2005]~287
which nicely coincides within the 3$\sigma$ positional error of 3.6\arcsec\ 
with the nova position. The source was not present in observation c3 half 
a year earlier. Also for the \chandra\ ACIS~S observation 1575 37~d after outburst
we can only determine an upper limit which however is not very constraining.
During the HRC~I observation 1912 63~d after the outburst we detect 
a source at the nova position with ($8.1\pm3.5$) counts. It will be interesting 
to see if the X-ray brightness of this well sampled fast
nova further increases.

\subsection{Nova in \m31\ [CFN87]~2}\label{[CFN87]2}
This nova was reported by CFN87 from eight H$\alpha$ images spread over 12~d.
The proposed X-ray counterpart [PFJ]~33 closely coincides with the nova position.
The source can not be resolved from nearby bright sources with the ROSAT PSPC in survey
one and survey two and was no longer detected in ROSAT HRI observations more than 
4 yrs later.
The X-ray source was detected  7.9 yrs after the nova outburst. This
could indicate that nova [CFN87]~2 is a recurrent nova which had a 
new outburst after about 7 years that was not optically detected and that is 
responsible for the observed X-rays.
As only the HRI detected the source we have no information on the X-ray spectrum.

\subsection{Nova in \m31\ [SI2001]~1990-08}\label{1990-08}
This nova was reported by SI2001 from two H$\alpha$ images 44~d apart.
The position is estimated by the authors to be accurate to $\sim$1\arcsec.  
The proposed X-ray counterpart [SHP97]~181 has a strong soft component and
the novae position is well within the 3$\sigma$ X-ray error radius. A hard
source was detected in the second ROSAT survey of \m31\ ([SHL2001]~181) 
19\arcsec\ offset to the NE. As this source was brighter and the sources
coincided within the 3$\sigma$ error circles of 15\arcsec, SHL2001 assumed that the
two sources from the first and second ROSAT survey were the same with the 
position of [SHL2001]~181. Inspection of the ROSAT
images clearly shows that the source in the second ROSAT survey is not the same 
source as the one detected in survey one. Due to the proximity of the new
source in survey two, no useful upper limits can be determined during these 
epochs. The source was no longer active in ROSAT HRI observations 3 yrs after 
the PSPC detections. 

\subsection{Nova in \m31\ [SI2001]~1997-06}\label{1997-06}
This nova was detected by SI2001 in two H$\alpha$ images 30~d apart.
The nova is also reported by RJC99 from two images, the first already 
exposed 44~d before
the first detection by SI2001. We therefore adopt nova position and date of 
outburst from  RJC99.
In X-rays, a source compatible with the nova position was detected about 
2.5 years after outburst during the
first \chandra\ HRC~I observation to the center of \m31. 
The source is no longer detected in the \chandra\ 
ACIS~S observation 1575, about 650~d later. The position of the 
source only $\sim$2\arcsec\ from the bright hard source [PFH2005]~299
prevents the detection with \xmm\ EPIC. The \chandra\ HRC~I provides no spectral 
information and there is no ACIS~S detection to identify it as a SSS.

\subsection{Nova in \m31\ AGPV~1576} \label{AGPV1576}
This source was detected by \citet{2004A&A...421..509A} from the AGAPE
project
and noted as nova candidate (already end of September 1994) 
from the pixellensing test project by TC96.
It also is a nova candidate from RJC99 (still visible in H$\alpha$ in
September 1995). 

The source was first detected in X-rays in the \chandra\ HRC~I observation
1912 (K2002) more than 7 yrs after outburst (only upper limit 124~d earlier
in \xmm\ observation c3). In the \xmm\ observation
67 days later the source can be classified as SSS. The novae position is well within 
the 3$\sigma$ X-ray error radii. The nova is not in the FOV of the 
ACIS~S observation 1575.

Also in this case, the rise of the X-ray flux of the nova rather 
late after the optical outburst 
could indicate that nova AGPV~1576 is a recurrent nova which had an 
new outburst after about 7 years that was not optically detected and that is 
responsible for the observed X-rays.

\subsection{Nova in \m31\ Sep-95}\label{Sep-95} 
This nova was reported by RJC99 from one H$\alpha$ image taken September 3,
1995. \chandra\ observations of K2002 and DKG2004 report an X-ray source at the 
nova position well within the positional errors which can be classified as SSS
using ACIS~S. The source position in the diffuse emission close to the \m31\
center near source [PFH2005]~310 prevented the detection by the \xmm\ EPIC 
instruments. The source is detected beginning with the first \chandra\ ACIS~S 
observations in June/July 2000 2.8 yrs after outburst. It is not visible 
in the first HRC~I observation half a year earlier. However, the 3$\sigma$ 
upper limit derived from this 5.2~ks observation is not much lower than 
the luminosity when the source is detected.  After  
\xmm\ observation c4, in each of the 1~ks \chandra\ HRC~I observations  3 to 5 counts 
are detected from the nova position
indicating that the SSS was still active 3100~d after outburst.

\subsection{Nova in \m31\ WeCAPP-N2000-03}\label{WeCAPP-N2000-03}
\begin{figure}
   \resizebox{\hsize}{!}{\includegraphics[angle=-90,clip]{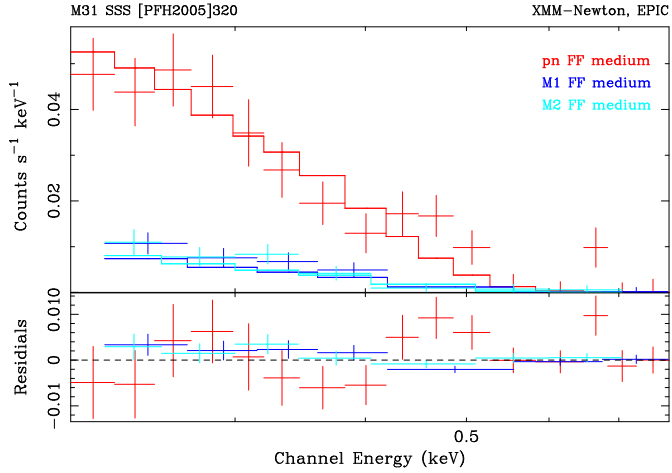}}
     \caption[]{\xmm\ EPIC spectrum of source [PFH2005]~320 
     (nova in \m31\ WeCAPP-N2000-03) for
     observation c3. The absorbed black body fit
     to the data (see Sect.~\ref{WeCAPP-N2000-03}) is shown in the upper panel. 
     }
    \label{spec_320} 
\end{figure}
Nova WeCAPP-N2000-03 was detected in outburst after a 18 d observing gap
and can be classified as fast nova. 
The POINT-AGAPE coverage \citep[nova PACN-00-01,][]{2004MNRAS.353..571D}
of the nova outburst starts a few days after the WeCAPP light curve
(see Fig.~\ref{nova_opt}). Brightness estimates from  
\citet[][]{2000IAUC.7477....2P} found the outburst maximum about 1.5 d before
the start of the dense WeCAPP monitoring.

In X-rays, the source ([PFH2005]~320) was first detected in the \chandra\ ACIS~S
observation 1854 170~d after outburst. In \xmm\ observation c2 16 days earlier
it was not detected (at least a factor of 5 fainter). It is classified as a SSS 
by PFH2005 and DKG2004. The source
stayed bright during the following \chandra\ HRC~I and ACIS~S
and the last \xmm\ observation until at least 528 days after outburst. 
EPIC spectra were extracted for this SSS 
and simultaneously fit with a black body 
model with \nh\ fixed to the foreground value. The best fit 
resulted in a black body temperature of ($33\pm3$) eV for observation c3. 
The spectra together with the 
best fit are shown in Fig.~\ref{spec_320}. 
During observation c4 the temperature was derived to (31$\pm$3) eV, consistent within
the errors to observation c3. The inferred unabsorbed luminosities 
(0.2--1.0 keV) during both observations are 3.1\ergs{37} and 2.2\ergs{37},
respectively. The spectral fits show that the luminosities given in 
Table~\ref{novae} for this source are underestimated by about a factor of 3.5. 
This is caused by the lower temperature 
of $\sim$30 eV while for the table  k$T$ = 50 eV was assumed. After  
\xmm\ observation c4, in each of the 1~ks \chandra\ HRC~I observations  2 to 3 counts are detected
from the nova position indicating that the SSS was still active 675~d after outburst.
It will be interesting to follow the X-ray light curve of this 
nova with accurately defined outburst epoch that has also been well sampled in X-rays
from the start of the outburst. 

\subsection{Nova in \m31\ WeCAPP-N2000-05}\label{WeCAPP-N2000-05}
Nova WeCAPP-N2000-05 was detected in outburst at the beginning of the WeCAPP
monitoring and showed a second maximum after about 42 days.
It is probably a dwarf nova with fast decline.
\chandra\ observations of K2002 and DKG2004 report an X-ray source at the 
nova position well within the positional errors which can be classified as SSS
using ACIS~S. Due to the bright source [PFH2005]~341 within 18\arcsec\ 
the nova is not resolvable by \xmm\ EPIC. While the source is already active
during \chandra\ observation 1570 352~d after outburst, no significant emission
was detected during the ACIS~S observation 1854 203~d after outburst.

\subsection{Nova in \m31\ [SI2001]~1997-09}\label{1997-09}
This nova was reported by SI2001 from four H$\alpha$ images 93~d apart.
The nova is also reported by RJC99. 
In X-rays, a SSS was first detected compatible with the nova position during 
\xmm\ observation c4 more than 4 years after 
outburst. It is not detected 67~d before in the \chandra\ HRC~I observation 1912
and 191~d before in the \xmm\ observation c3, respectively. 
Due to the rather large off-axis angle (8.5\arcmin), the \chandra\ upper limits are
rather large. The nova is not in the FOV of the \chandra\ ACIS~S observation 1575.
 
\subsection{Nova in \m31\ GCVS-M31-V0962}\label{GCVS-M31-V0962}
This nova was reported by SI2001 as 1996-05 from one H$\alpha$ image.
It coincides in position with [H29]~N40 \citep{1929ApJ....69..103H} and
is therefore classified as recurrent nova ($\Delta T \sim 72$ yr or shorter).
In X-rays, a source compatible with the position of the nova was reported
by K2002 in HRC~I observation 1912, 1906 days after outburst. 
In the \xmm\ observation 67 days later it is identified with the 
SSS [PFH2005]~359. The source position is at the boarder of ACIS~S3
CCD in \chandra\ observation 1575 and not in the source list of DKG2004.
However, we determined a source luminosity during this observation 
which is compatible with that of the 
\chandra\ HRC~I observation 26~d later. In the \xmm\ observation c3
98~d earlier and the \xmm\ observations before the source is not
detected.

\subsection{Nova in \m31\ [SI2001]~1995-05}\label{1995-05}
This nova was reported by SI2001 from two H$\alpha$ images 47~d apart.
In X-rays, a source compatible with the position of the nova was detected as SSS
in each of the four \xmm\ observations to the \m31\ center and in the
\chandra\ observations of K2002 and DKG2004,  4.5 to 6.1 yrs after outburst.
During this time the X-ray luminosity was rising by about a factor of
three. The \chandra\ luminosities between \xmm\ observation c3 and c4 
seem to indicate even higher source luminosity. However, it is more likely that
this difference reflects ECF uncertainties. This explanation is supported by
the \chandra\ luminosities derived from the short earlier observations
starting from 4 yrs after outburst. In a ROSAT HRI observation 37~d after 
the nova outburst the source was not detected. 
Note that the long duration of the SSS emission in this source is only 
comparable to the galactic classical nova GQ Mus 
\citep{1993Natur.361..331O} or to Nova LMC 1995 \citep{2003ApJ...594..435O}.

\subsection{Nova in \m31\ [SI2001]~1992-04} \label{1992-04}
This nova was reported by SI2001 from two H$\alpha$ images 35~d apart. The
proposed X-ray counterpart was detected in two ROSAT HRI observations about
570~d after outburst. Due to its position close to the \m31\ center it could
not be resolved in the ROSAT PSPC surveys. The upper limit before  
the detections originates from before the nova outburst. Also the upper 
limits after the detection do not constrain the length of the
X-ray on state. 
As only the HRI detected the source we have no information on the X-ray spectrum.

\subsection{Nova in \m31\ Jul-98}\label{Jul-98}
\begin{figure}
   \resizebox{\hsize}{!}{\includegraphics[angle=-90,clip]{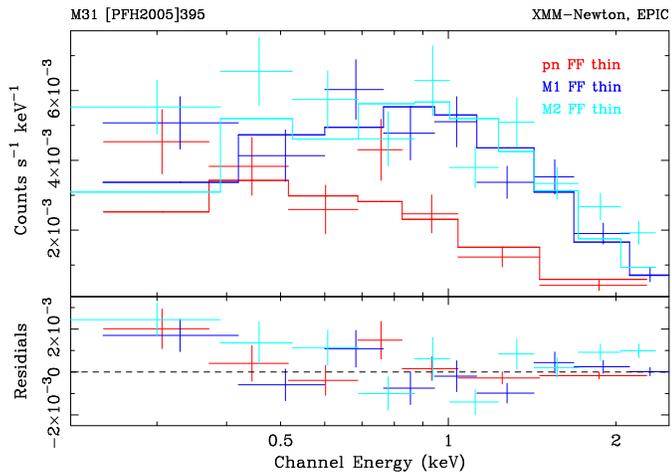}}
     \caption[]{\xmm\ EPIC spectrum of source [PFH2005]~395 
     (nova in \m31\ Jul-98) for
     observation c4. The absorbed bremsstrahlung fit
     to the data (see Sect.~\ref{Jul-98}) is shown in the upper panel. The reduced 
     flux in the pn spectrum is due to the location of the source on a CCD gap.
     }
    \label{spec_395} 
\end{figure}
This nova was reported by RJC99 from H$\alpha$ images in June 1998 to July 1999. It
was brightest on an H$\alpha$ image exposed on July 25, 1998.
In X-rays, the source was first detected in \xmm\ observation c3 about
three years
after outburst and brightened to the following \chandra\ HRC~I and ACIS~S
and the \xmm\ observation c4. In a 1.1~ks HRC~I observation 10~d after observation c4
it was brighter by another factor of 4. In a 1.2~ks HRC~I observation 150~d after
observation c4, the source was not detected and the upper limit indicates a 
significant decrease in brightness.
The nova was not detected in \xmm\ observation c1 and c2, 369~d and 183~d before 
observation c3, respectively.
PFH2005 classified the source as candidate for an X-ray binary due
to its transient behavior and its hard spectrum. 
EPIC MOS and pn spectra extracted for observation c4
extend to energies above 
2.5 keV and confirm that [PFH2005]~395 during this observation is not a SSS. 
Simultaneous fits -- with free normalization for pn as the source in this
instrument is partly
located on a CCD gap -- yield 
an unacceptable reduced $\chi^2$ of 3.3  for 65 degrees of freedom for an 
absorbed black body model with a strong low-energy excess.  A bremsstrahlung model 
with a temperature ($1.34^{+0.29}_{-0.23}$)~keV (\nh\ fixed at Galactic 
foreground) represents the spectra best ($\chi^2$ = 106, 65 dof) among 
simple one-component models. The spectra together with the best fit are 
shown in Fig.~\ref{spec_395}. The unabsorbed luminosities inferred from the MOS
spectra in
the (0.2--1.0) keV and (0.2--2.0) keV bands are 5.3\ergs{36} and 7.8\ergs{36},
respectively. The spectral fits show that the luminosities given in 
Table~\ref{novae} for this source are overestimated by about a factor of four
due to the wrong spectral model assumed. If the X-ray spectrum stayed the same, 
the flux for \chandra\ HRC I observation 2905 is overestimated by a factor of
about nine.

During \xmm\ observation c3 when the source was still faint, hardness ratios 
are typical for a SSS. 
Also DKG2004 classify the source as SSS based on \chandra\ ACIS~S 
observation 1575 about 100~d later.
Using HST images DKG2004 detected a star within the 1\arcsec\ error circle of the \chandra\ 
position with brightness and colors compatible with a symbiotic. This tentative 
identification was based on the fact that several of the SSSs in the Galaxy and the 
Magellanic Clouds are symbiotics. However only few symbiotics show the very soft X-ray 
spectrum of a SSS with emission mainly below 0.5 keV. Another group of symbiotics exhibits 
X-ray spectra with the emission peaking around 0.8 keV and might be explained 
by emission 
from optically thin plasma with temperatures in the range of a few 10$^6$ K to a few 10$^7$ K 
caused by colliding winds \citep{1997A&A...319..201M} in the binary system. 
The EPIC spectra of [PFH2005]~395 are compatible with such a model which supports the 
identification as symbiotic.
However, the distinct spectral change from very soft (0.5 keV) to a few keV
spectrum on a timescale of a year has not been seen in symbiotics before,
so this interpretation might not be true. An alternative might be
a behaviour seen in the source 1E1339.8+2837 in M~3 \citep{1999PASJ...51..519D} 
which was found to switch between
supersoft and hard states on timescales of 6 months. The main difference,
however, are the luminosities: this source had only $10^{35}$ erg s$^{-1}$
in the soft state, and about $10^{33}$ erg s$^{-1}$ in the hard state.
Yet another alternative are black hole transients, which usually also
make soft to hard transitions, but also with luminosity changes. In the 
latter cases the source would not be an optical nova. The soft/hard spectral 
transition with correlated luminosity changes makes the source unique in our 
sample and may indicate that the
H$\alpha$ outburst of the source was incorrectly classified as optical nova.

\subsection{Nova in \m31\ [SI2001]~1990-12}\label{1990-12}
This nova was reported by SI2001 from two H$\alpha$ images 45~d apart.
The proposed X-ray counterpart [SHL2001]~230 has a strong soft component and
the novae position is well within the 3$\sigma$ X-ray error radius.
The X-ray source was bright during SII-E1 and SII-E2 605~d and 755~d after
outburst. In SII-E3 (955~d) the luminosity 
decreased by a factor of two. During the first
survey (229~d), the source was not detected. However the upper limit is rather high  
as the source was always observed at rather large off-axis angles ($>$10\arcmin). 
The source was not detected in ROSAT HRI observations before and after the 
PSPC observations. Especially the upper limits of the HRI observations after the
PSPC detections are only mildly restraining the outburst duration.

\subsection{Nova in \m31\ WeCAPP-N2001-08}\label{WeCAPP-N2001-08}
Nova WeCAPP-N2001-08 was detected in outburst after a 30 d observing gap
and can be classified as fast nova. 
In X-rays, a source compatible with the nova position was reported from 
by K2002 from \chandra\ observation 2906 119 d after outburst. 
\xmm\ observations 124~d before and 67~d after the \chandra\ 
observation do not detect the source.
The nova position is outside the FOV of ACIS~S observation 1575.
As only the HRC~I detected the source we have no information of the
shape of the X-ray spectrum.

\subsection{Nova in \m31\ GCVS-M31-V1067}\label{GCVS-M31-V1067}
This nova was reported by RJC99 from H$\alpha$ images in June 1998 to July 1999. It
was brightest on an H$\alpha$ image exposed on June 6, 1998.
It coincides in position with [H29]~N86 \citep{1929ApJ....69..103H} and
is therefore classified as recurrent nova ($\Delta T \sim 79$ yr or shorter).
In X-rays, a source compatible with the position of the nova was  
detected as faint SSS in the \xmm\ observations c3 and c4 ([PFH2005]~456), 
1119 d and 1310 d after outburst. 
In the \xmm\ observations c1 and c2 936~d and 750~d after the nova outburst, 
respectively, the source was not detected. With its distance of $\sim$8.3\arcmin\
from the center of \m31\ the nova was not detected in the \chandra\ HRC~I
observation 1912 (3$\sigma$ upper limit well above the EPIC detection). The
nova was not in the FOV in \chandra\ ACIS~S observation 1575.

\subsection{Nova in \m31\ [SI2001]~1991-12}\label{1991-12}
This nova was reported by SI2001 from one H$\alpha$ image.
The proposed X-ray counterpart [SHL2001]~246 has a strong soft component and
the novae position is well within the 3$\sigma$ X-ray error radius.
The X-ray source was bright during SII-E1 560~d after
outburst and fainter in SII-E2 (710~d). During SII-E3 (910~d) and the first
survey (185~d), the source was not detected. The source was not detected in 
ROSAT HRI observations before and after the outburst. We do not give HRI upper 
limits for this source in Table~\ref{novae} as the PSPC upper limits are more 
constraining both in terms of luminosity and outburst duration.

\subsection{Nova in \m31\ [SI2001]~1990-16}\label{1990-16}
This nova was reported by SI2001 from two H$\alpha$ images 2~d apart.
The ROSAT and optical observations of the nova have been reported in detail by 
\citet{2002A&A...389..439N}. However, in the paper they only give an average 
upper limit for the full second survey. We derived upper limits for SII-E1 
and SII-E2 689~d and 839~d after outburst, respectively. Only one observation
of 1.8~ks was available during SII-E3 which did not give a sensitive
upper limit. The source was not detected during the ROSAT HRI observation 
150006h 52~d before the nova outburst. Due to the high off-axis position the
3$\sigma$ upper limit is not very constraining.

\subsection{Nova in \m31\ ShAl~57}\label{ShAl57}
A special case is
Nova ShAl~57 which correlates (well within the 3$\sigma$ X-ray error
radius) with the source [SHP97]~319 from the first ROSAT survey, which has a
strong soft component. The source vanished in 
ROSAT survey II. The optical outburst 
occured about 6 years after the ROSAT detection \citep{1998AstL...24..641S}. 
However, the nova may still be the correct identification if 
it is a recurrent nova. Then, a previous outburst responsible for the X-ray emission, 
might have been missed. The nova is not in the FOV of ROSAT HRI observations. 

\subsection{Nova in \me33\ [WS2004]~1995-1}\label{1995-1}
This nova was reported by \citet[][hereafter WS2004]{2004ApJ...612..867W}.
In X-rays, a source compatible with the position of the nova was 
reported in the \me33\ ROSAT source catalogue of HP2001 from the combined HRI
observations. We analyzed four ROSAT HRI observations that were separated by 
half a year with the first 44 days before the optical nova detection. Only in
the observation 145 d after the optical nova detection we find significant
flux. Half a year later the flux dropped by at least a factor of four.
As only the HRI detected the source we have no information on the X-ray spectrum.

\subsection{Nova in \me33\ [WS2004]~1995-3}\label{1995-3}
This nova was reported by WS2004.
In X-rays, a source compatible with the position of the nova is clearly detected 
in \chandra\ ACIS~S observation 786 about 5 years after the outburst. 
In this 46.3 ks observation to the \me33\ center, we detect 
$21.9\pm5.0$ and $2.9\pm3.5$  counts in the energy bands below 0.7 keV and 
(0.7--8.0) keV, respectively, identifying the source as supersoft. The PMH2004 
\xmm\ catalogue shows no source at the position. We searched the individual 
observations and detected a $>2\sigma$ excess at the source position in EPIC 
pn in observation 0102640101 which corresponds to a luminosity compatible with
that of the \chandra\ observation 26 days later. The source is not in the FOV 
of the MOS cameras during this observation. While the upper limits from a \xmm\ 
observation about one year later do not constrain the light curve, observations
6.3 yrs after the outburst seem to indicate a decline.  
The nova is not detected in several 
ROSAT HRI observations up to 1.2 yrs after the outburst. However, a source at the
\chandra\ luminosity would have escaped detection. 

\section{Discussion}
The \xmm\ catalogue of \m31\ lists 18 SSS (PFH2005). Seven coincide with optical 
nova positions. Another seven sources are not in the region, covered by
the WeCAPP survey. From the remaining four sources two ([PFH2005] 430 and 431)
are very bright in at
least one \xmm\ observation and [PFH2005] 431 shows a 865~s period in the \xmm\ data
reminiscent of a rotation period of a magnetized white dwarf 
\citep{2001A&A...378..800O}.
These findings may classify the latter two as SSS as known from the Magellanic
Clouds. These steady burning WDs in close binaries
\citep{1992A&A...262...97V} 
differ from novae by (I) either constant or on/off X-ray emission
and (II) by optical variations of at most one magnitude.
The nova searches in the center area of \m31\ are -- due to lacks in the 
optical coverage -- by far not complete and even more so in the outer disk
areas. Therefore most of the remaining unidentified SSS
could present X-ray emission from optical novae in the plateau phase.
This view is supported by the fact that all but one (the symbiotic nova 
candidate RJC99 Jul-98) of the nova correlations detected with \xmm\ and 
\chandra\ ACIS~S are classified as SSS. Also all of the ROSAT PSPC sources
correlating with novae, are mainly radiating in the band below 0.4 keV
and therefore SSS candidates.

The \xmm\ catalogue of \me33\ lists five SSS (PMH2004). None of them coincides with 
a known optical nova. However, this is not surprising and may just be caused by
low number statistics and still most of these SSS may represent nova in the supersoft 
X-ray phase after outburst. 
WS2004 estimate a global nova rate for \me33 of 2.5 yr$^{-1}$. However, a higher
nova rate of 4.6 novae per year as derived by \citet{1994A&A...287..403D}
from a relatively frequent B monitoring of the galaxy may be more realistic 
\citep{2004AJ....127..816N}.
If we assume that
novae from more than the past six years may show up in X-rays as is indicated 
from our correlations in Table~\ref{novae}, and take into account that the \xmm\ 
\me33\ survey was accumulated over 2 years, 20--40 novae with outburst dates 
from 1995 to end of 2002 could have been contributing. On the other hand, during this time scale just
six novae are reported in the literature (see WS2004), three with outburst in 1995,
one each in  1996, 1997 and 2001. In the years before the \xmm\ observations just
one nova was reported. Therefore most if not all of the novae radiating
in X-rays may have been missed in the catalogues. The detection of the two nova in \me33\
with the ROSAT HRI and \chandra\ may reflect the denser sampling in 1995. Therefore, if one wants 
to detect and identify optical novae as SSS in \me33\ in the future an efficient 
search program for optical novae will first be necessary.  

\subsection{Inference to close binary supersoft models}
We find that a major fraction of SSS as previously found with ROSAT, \xmm\ and \chandra\ 
are optical novae. This poses an interesting problem for the close-binary supersoft 
sources (CBSS) which
are expected to be a numerous class of object \citep{1994ApJ...437..733D}, with
$\sim$1000 expected for \m31. Since the Magellanic Clouds contain about a dozen CBSS,
the new statistics of SSS in \m31\ as presented here as remainders beyond the nova 
population indicate that the number of SSS seemingly does not scale with the mass 
of the host galaxy. It thus may be worth-while to re-consider the population studies
and the detectability of CBSS.

\subsection{Nova parameters}
The bolometric luminosity of a white dwarf envelope with hydrogen
burning is directly related to the white dwarf mass 
\citep{1998ApJ...503..381T,SH2005a}.  Unfortunately, the bolometric luminosity
of the SSS sources detected can not be well determined with the present
observations. Nevertheless, the evolution of the effective temperature of
the source, which could be only obtained with well sampled X-ray
monitoring observations, can also constrain the parameters of the white
dwarf, independently of the luminosity \citep[as done recently for V1974 Cyg by][]{SH2005b}.

Assuming that the material ejected by
the nova explosion forms a spherical, homogeneous shell expanding at
constant velocity $v$, the hydrogen mass density of the shell will evolve in
time $t$ like $\rho =\frac{M_{H}}{\frac{4}{3}\pi v^{3}t^{3}}$ where M$_{H}$
is the ejected hydrogen mass \citep{1996ApJ...456..788K}. Assuming a constant
density, the column density of hydrogen will evolve with time like
$\nh({\rm cm}^{-2})=\frac{M_{H}}{\frac{4}{3}\pi m_{H}v^{2}t^{2}}$, where
$m_{H}$ is the mass of the hydrogen atom.  Assuming typical values for the
expansion velocity (1000 km/s) and the ejected hydrogen mass
($10^{-5}$~M$_{\odot}$), the absorption column should reduce to $\sim$
\ohcm{21} and thus be transparent to soft X-rays in less than one
year.  Indeed, for the known classical novae with soft X-ray emission, the
turn on always occurred within the first year after the outburst 
\citep[see for instance][]{2004RMxAC..20..182O}.
We can use the well sampled X-ray light curves of novae WeCAPP-N2001-12
(turn on of SSS state between days 37 and 63 after outburst) and WeCAPP-N2000-03 
(days 154 to 170) to estimate -- under the assumptions above 
-- an ejected hydrogen mass of $\sim10^{-6}M_{\odot}$ and  
$\sim10^{-5}M_{\odot}$ in the corresponding nova outbursts, respectively. 

We can compare these results with the ejected mass derived from the 
relationship between the hydrogen ejected mass and the rate of decline 
found by \citet{2002A&A...390..155D}, $log M^{ej}_{H}(10^{-5} M_{\odot})= 
0.274(\pm0.197)\times log t_{2}+0.645(\pm0.283)$. For Nova WeCAPP-N2000-03, 
the outburst is nearly fully covered in the optical (see discussion in
Sect.~\ref{WeCAPP-N2000-03}), and its $t_{2}$ was $\sim 7.3$ days in R and $\sim 
9.7$ days in I, which with the above relation indicates an hydrogen 
ejected mass in the range (taking into account the error bars) 
$(3-20)\times 10^{-5}M_{\odot}$, larger than the mass derived from the 
X-ray light-curve. For Nova WeCAPP-N2001-12, we have only an upper limit 
for $t_{2}$ of 11 days, leading to an upper limit for the ejected mass of 
$3\times 10^{-4}M_{\odot}$. In any case, the ejected masses derived both 
from the X-ray and the optical light-curves are hydrogen masses, and 
therefore lower limits to the total ejected mass.

\subsection{X-ray detection of optical novae}
\begin{figure}
   \resizebox{\hsize}{!}{\includegraphics[angle=-90,clip]{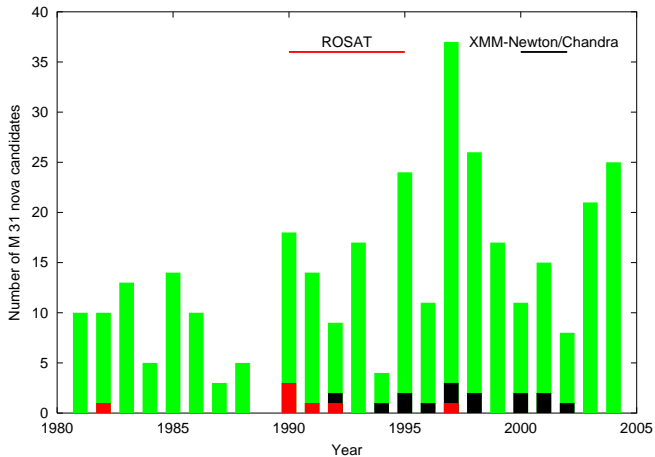}}
     \caption[]{Histogram of the number of optical novae per year in \m31
     contained in the nova catalogue used for X-ray cross-correlation (see 
     Sect.~\ref{optnova}). The number of optical novae showing X-ray emission 
     is indicated coded for ROSAT and \xmm/\chandra. The time span of X-ray
     observations is indicated.
     }
    \label{nova_year} 
\end{figure}
Figure~\ref{nova_year} shows the number of optical novae per year in the optical nova 
catalogue of \m31 used for cross-correlation with the X-ray data. X-ray detected 
novae are indicated separating ROSAT and \xmm/\chandra\ detections. The time span 
of the \m31\ observations of these satellites is also indicated. There are only few
optical novae detected in the years before the ROSAT observations which may explain 
the lack of nova detections before the ROSAT observations. With \xmm\ and \chandra\ 
many novae are detected which had their outburst years before the X-ray outburst.
As described for the individual novae and discussed in more detail below, this may
be caused due to long supersoft states of these novae or indicate that some of these
novae are recurrent and had a new outburst short before the X-ray observations.
Four of the 25 optical novae in 2000 and 2001 (16\%) are detected to turn on
within a year. This percentage is only a lower limit as some novae with short supersoft
states (shorter than 6 month) may have been missed in the sparse and inhomogeneous 
sampling of the light curves and novae in the crowded center area of \m31\ are missed 
in \xmm\ observations.
Only prolonged sampling will allow us to decide how many additional
novae will show a late turn-on as an X-ray source.

In seven of the sources
the soft X-ray emission is observed to turn on during the first year after
the outburst, in three the emission is already detected in the first X-ray observation.
Eleven of the detected novae have been observed to turn-off, 
and in five cases the SSS is still bright 5 yrs after
outburst (Novae in M31 [SI2001] 1992-01, 1995-05, AGPV 1576, RJC99 Sep-95
and GCVS-M31-V0962). Both the large
fraction of SSS detected among optical novae candidates and the long
duration of the soft X-ray emission support the expected presence of a
post-outburst hydrogen burning envelope left on the white dwarf.

\subsection{Supersoft X-ray emission from recurrent novae?}
Surprisingly enough, six of the SSS detected have turned on only between
3 and 9 years after the optically discovered nova outburst.  This
``delayed" appearance of the SSS is unlikely to be associated to the
decrease of absorption in the ejecta.

A change in the accretion rate after the nova outburst up to the level of
powering a SSS with steady hydrogen burning also seems unlikely. 
\citet{1988ApJ...325..828K}
showed that the irradiation of the secondary star after the
nova outburst could cause the red dwarf to expand and induce a mass
transfer rate enhanced by two orders of magnitude. They assumed that the
luminosity of the irradiation source (the hydrogen burning envelope on the
white dwarf) was constant and of the order of the Eddington luminosity for
a certain time and that, after the irradiating source turned off, the
white dwarf started to cool down. The maximum radii of the secondary star
was achieved shortly ($\sim$ 0.1 yr) after the end of the constant
luminosity phase, and then started to slowly contract, thus decreasing the
accretion rate.  If this process had occurred in the delayed SSS detected
in M31, the irradiating white dwarf with a luminosity close to the
Eddington limit would have been visible in X-rays as soon as the ejected
shell became optically thin to X-rays (as mentioned above, within the
first year after the outburst). But for these five cases, no SSS was
detected in the first observations, performed between $\sim 700$ and $\sim
3000$ days after the outburst, depending on the source.

Finally, it is possible that these five nova candidates with delayed X-ray
emission are recurrent novae, and that the optical outburst responsible
for the X-ray emission was not detected. The same would be true for the
nova candidate which showed X-ray emission 6 yrs before the nova outburst
(ShAl 57). Two of the five delayed SSS are
in fact classified as recurrent novae. If all these nova candidates 
are indeed recurrent novae, 29\% of the novae
detected as SSSs would be recurrent novae, a rate compatible with the 
upper limit estimated by \citet{1996ApJ...473..240D} for \m31. Detection of
delayed SSS states in X-rays may in the future be used as a new method to 
classify recurrent novae in \m31 and to derive the ratio of recurrent novae to 
classical novae. 

\section{Summary and conclusions}
We searched for X-ray counterparts of optical novae detected in \m31\ and \me33.  
We combined an optical nova catalogue from the WeCAPP survey with optical novae
reported in the literature and correlated them with the most recent X-ray catalogues
from ROSAT, \xmm\ and \chandra, and -- in addition -- searched for nova correlations 
in archival data. We report 21 X-ray counterparts for novae in \m31 (mostly SSS) and
two in \me33. Our sample more than triples the number of known optical novae with supersoft phase. 
Most of the counterparts are covered in several observations allowing 
us to constrain their X-ray light curves. Selected brighter sources were
classified by their \xmm\ EPIC spectra. Six counterparts are only detected in
\chandra\ HRC~I (3) or ROSAT HRI (3) observations, i.e. X-ray detectors with no
energy resolution, and therefore can not be classified as supersoft.

We estimated the fraction of novae turning on as SSS within a year.
The X-ray observations can be used to constrain the parameters of the white dwarf 
and determine the mass of the material ejected in the outburst. 

More information can be expected from the analysis of already performed 
additional \xmm\ and \chandra\ observations to the \m31\ center that are 
not yet public. However, these observations are not homogeneously sampling 
the expected soft X-ray nova light curves and a dedicated observation campaign
for \m31\ novae would be desirable. Such a campaign should cover several years 
in optical and X-rays and also allow to constrain the X-ray light curve of 
novae that are only X-ray visible for a few months. 

Ongoing optical and X-ray monitoring of the central region of \m31\, where most
of the novae are detected, should allow us to determine the length of the
plateau phase of several novae and together with nova temperature development
give a handle on the masses of the white dwarf involved. Due to the
simultaneous X-ray coverage of several novae at a time and the known distance
of the novae, such a program promises insights in the nova phaenomenon that
are much more difficult to obtain from the observation of novae in the Milky
Way.
 
\begin{acknowledgements}
We thank the referee Massimo Della Valle for his comments, which helped
to improve the manuscript considerably.
Part of this work was supported by the 
\emph{Son\-der\-for\-schungs\-be\-reich, SFB\/} 375
of the \emph{Deut\-sche For\-schungs\-ge\-mein\-schaft, DFG\/}.
The \xmm\ project is supported by the Bundesministerium f\"{u}r
Bildung und Forschung / Deutsches Zentrum f\"{u}r Luft- und Raumfahrt 
(BMBF/DLR), the Max-Planck Society and the Heidenhain-Stiftung.
\end{acknowledgements}

\bibliographystyle{aa}
\bibliography{./3127,/home/wnp/data1/papers/my1990,/home/wnp/data1/papers/my2000,/home/wnp/data1/papers/my2001}

\begin{thebibliography}{84}
\expandafter\ifx\csname natexlab\endcsname\relax\def\natexlab#1{#1}\fi

\bibitem[{{Alard} \& {Lupton}(1998)}]{1998ApJ...503..325A}
{Alard}, C. \& {Lupton}, R.~H. 1998, \apj, 503, 325

\bibitem[{{An} {et~al.}(2004){An}, {Evans}, {Hewett}, {Baillon}, {Calchi
  Novati}, {Carr}, {Cr{\' e}z{\' e}}, {Giraud-H{\' e}raud}, {Gould}, {Jetzer},
  {Kaplan}, {Kerins}, {Paulin-Henriksson}, {Smartt}, {Stalin}, \&
  {Tsapras}}]{2004MNRAS.351.1071A}
{An}, J.~H., {Evans}, N.~W., {Hewett}, P., {et~al.} 2004, \mnras, 351, 1071

\bibitem[{{Ansari} {et~al.}(2004){Ansari}, {Auri{\` e}re}, {Baillon},
  {Bouquet}, {Coupinot}, {Coutures}, {Ghesqui{\` e}re}, {Giraud-H{\' e}raud},
  {Gillieron}, {Gondolo}, {Hecquet}, {Kaplan}, {Kim}, {Le Du}, {Melchior},
  {Moniez}, {Picat}, \& {Soucail}}]{2004A&A...421..509A}
{Ansari}, R., {Auri{\` e}re}, M., {Baillon}, P., {et~al.} 2004, \aap, 421, 509

\bibitem[{{Burwitz} {et~al.}(2002){Burwitz}, {Starrfield}, {Krautter}, \&
  {Ness}}]{2002AIPC..637..377B}
{Burwitz}, V., {Starrfield}, S., {Krautter}, J., \& {Ness}, J. 2002, in AIP
  Conf. Proc. 637: Classical Nova Explosions, eds. M. Hernanz \& J. Jos\'e,
  p.377

\bibitem[{{Ciardullo} {et~al.}(1987){Ciardullo}, {Ford}, {Neill}, {Jacoby}, \&
  {Shafter}}]{1987ApJ...318..520C}
{Ciardullo}, R., {Ford}, H.~C., {Neill}, J.~D., {Jacoby}, G.~H., \& {Shafter},
  A.~W. 1987, \apj, 318, 520

\bibitem[{{Ciardullo} {et~al.}(1990){Ciardullo}, {Shafter}, {Ford}, {Neill},
  {Shara}, \& {Tomaney}}]{1990ApJ...356..472C}
{Ciardullo}, R., {Shafter}, A.~W., {Ford}, H.~C., {et~al.} 1990, \apj, 356, 472

\bibitem[{{Darnley} {et~al.}(2004){Darnley}, {Bode}, {Kerins}, {Newsam}, {An},
  {Baillon}, {Novati}, {Carr}, {Cr{\' e}z{\' e}}, {Evans}, {Giraud-H{\'
  e}raud}, {Gould}, {Hewett}, {Jetzer}, {Kaplan}, {Paulin-Henriksson},
  {Smartt}, {Stalin}, \& {Tsapras}}]{2004MNRAS.353..571D}
{Darnley}, M.~J., {Bode}, M.~F., {Kerins}, E., {et~al.} 2004, \mnras, 353, 571

\bibitem[{{Della Valle} \& {Livio}(1996)}]{1996ApJ...473..240D}
{Della Valle}, M. \& {Livio}, M. 1996, \apj, 473, 240

\bibitem[{{Della Valle} {et~al.}(2002){Della Valle}, {Pasquini}, {Daou}, \&
  {Williams}}]{2002A&A...390..155D}
{Della Valle}, M., {Pasquini}, L., {Daou}, D., \& {Williams}, R.~E. 2002, \aap,
  390, 155

\bibitem[{{Della Valle} {et~al.}(1994){Della Valle}, {Rosino}, {Bianchini}, \&
  {Livio}}]{1994A&A...287..403D}
{Della Valle}, M., {Rosino}, L., {Bianchini}, A., \& {Livio}, M. 1994, \aap,
  287, 403

\bibitem[{{Di Stefano} {et~al.}(2004){Di Stefano}, {Kong}, {Greiner},
  {Primini}, {Garcia}, {Barmby}, {Massey}, {Hodge}, {Williams}, {Murray},
  {Curry}, \& {Russo}}]{2004ApJ...610..247D}
{Di Stefano}, R., {Kong}, A.~K.~H., {Greiner}, J., {et~al.} 2004, \apj, 610,
  247

\bibitem[{{Di Stefano} \& {Rappaport}(1994)}]{1994ApJ...437..733D}
{Di Stefano}, R. \& {Rappaport}, S. 1994, \apj, 437, 733

\bibitem[{{Dotani} {et~al.}(1999){Dotani}, {Asai}, \&
  {Greiner}}]{1999PASJ...51..519D}
{Dotani}, T., {Asai}, K., \& {Greiner}, J. 1999, \pasj, 51, 519

\bibitem[{{Drake} {et~al.}(2003){Drake}, {Wagner}, {Starrfield}, {Butt},
  {Krautter}, {Bond}, {Della Valle}, {Gehrz}, {Woodward}, {Evans}, {Orio},
  {Hauschildt}, {Hernanz}, {Mukai}, \& {Truran}}]{2003ApJ...584..448D}
{Drake}, J.~J., {Wagner}, R.~M., {Starrfield}, S., {et~al.} 2003, \apj, 584,
  448

\bibitem[{{Fiaschi} {et~al.}(2001){Fiaschi}, {Di Mille}, \&
  {Cariolato}}]{2001IAUC.7709....3F}
{Fiaschi}, M., {Di Mille}, F., \& {Cariolato}, R. 2001, \iaucirc, 7709

\bibitem[{{Fiaschi} {et~al.}(2002){Fiaschi}, {Di Mille}, {Cariolato}, {Swift},
  \& {Li}}]{2002IAUC.7794....1F}
{Fiaschi}, M., {Di Mille}, F., {Cariolato}, R., {Swift}, B., \& {Li}, W.~D.
  2002, \iaucirc, 7794

\bibitem[{{Fliri} {et~al.}(2005){Fliri}, {Riffeser}, {Seitz}, \&
  {Bender}}]{wecapp04}
{Fliri}, J., {Riffeser}, A., {Seitz}, S., \& {Bender}, R. 2005, \aap, submitted

\bibitem[{{G{\" o}ssl} \& {Riffeser}(2002)}]{2002A&A...381.1095G}
{G{\" o}ssl}, C.~A. \& {Riffeser}, A. 2002, \aap, 381, 1095

\bibitem[{{Gonzalez-Riestra} {et~al.}(1998){Gonzalez-Riestra}, {Orio}, \&
  {Gallagher}}]{1998A&AS..129...23G}
{Gonzalez-Riestra}, R., {Orio}, M., \& {Gallagher}, J. 1998, \aaps, 129, 23

\bibitem[{{Greiner} {et~al.}(2004){Greiner}, {Di Stefano}, {Kong}, \&
  {Primini}}]{2004ApJ...610..261G}
{Greiner}, J., {Di Stefano}, R., {Kong}, A., \& {Primini}, F. 2004, \apj, 610,
  261

\bibitem[{{Greiner} {et~al.}(1996){Greiner}, {Supper}, \&
  {Magnier}}]{1996LNP96.472...75G}
{Greiner}, J., {Supper}, R., \& {Magnier}, E.~A. 1996, in Supersoft X-Ray
  Sources, Lecture Notes in Physics, ed. J. Greiner, Springer(Berlin Heidelberg
  New York), 472, p.75

\bibitem[{{Haberl} \& {Pietsch}(2001)}]{2001A&A...373..438H}
{Haberl}, F. \& {Pietsch}, W. 2001, \aap, 373, 438

\bibitem[{{Holland}(1998)}]{1998AJ....115.1916H}
{Holland}, S. 1998, \aj, 115, 1916

\bibitem[{{Hubble}(1929)}]{1929ApJ....69..103H}
{Hubble}, E.~P. 1929, \apj, 69, 103

\bibitem[{{Jansen} {et~al.}(2001){Jansen}, {Lumb}, {Altieri}, {Clavel}, {Ehle},
  {Erd}, {Gabriel}, {Guainazzi}, {Gondoin}, {Much}, {Munoz}, {Santos},
  {Schartel}, {Texier}, \& {Vacanti}}]{2001A&A...365L...1J}
{Jansen}, F., {Lumb}, D., {Altieri}, B., {et~al.} 2001, \aap, 365, L1

\bibitem[{{Johnson} {et~al.}(1999){Johnson}, {Modjaz}, \&
  {Li}}]{1999IAUC.7236....1J}
{Johnson}, R., {Modjaz}, M., \& {Li}, W.~D. 1999, \iaucirc, 7236

\bibitem[{{Joshi} {et~al.}(2004){Joshi}, {Pandey}, {Narasimha}, {Giraud-H{\'
  e}raud}, {Sagar}, \& {Kaplan}}]{2004A&A...415..471J}
{Joshi}, Y.~C., {Pandey}, A.~K., {Narasimha}, D., {et~al.} 2004, \aap, 415, 471

\bibitem[{{Kaaret}(2002)}]{2002ApJ...578..114K}
{Kaaret}, P. 2002, \apj, 578, 114

\bibitem[{{Kato} \& {Hachisu}(1994)}]{1994ApJ...437..802K}
{Kato}, M. \& {Hachisu}, I. 1994, \apj, 437, 802

\bibitem[{{Kong} {et~al.}(2002){Kong}, {Garcia}, {Primini}, {Murray}, {Di
  Stefano}, \& {McClintock}}]{2002ApJ...577..738K}
{Kong}, A.~K.~H., {Garcia}, M.~R., {Primini}, F.~A., {et~al.} 2002, \apj, 577,
  738

\bibitem[{{Kovetz} {et~al.}(1988){Kovetz}, {Prialnik}, \&
  {Shara}}]{1988ApJ...325..828K}
{Kovetz}, A., {Prialnik}, D., \& {Shara}, M.~M. 1988, \apj, 325, 828

\bibitem[{{Krautter} {et~al.}(1996){Krautter}, {\"Ogelman}, {Starrfield},
  {Wichmann}, \& {Pfeffermann}}]{1996ApJ...456..788K}
{Krautter}, J., {\"Ogelman}, H., {Starrfield}, S., {Wichmann}, R., \&
  {Pfeffermann}, E. 1996, \apj, 456, 788

\bibitem[{{MacDonald} {et~al.}(1985){MacDonald}, {Fujimoto}, \&
  {Truran}}]{1985ApJ...294..263M}
{MacDonald}, J., {Fujimoto}, M.~Y., \& {Truran}, J.~W. 1985, \apj, 294, 263

\bibitem[{{M\"urset} {et~al.}(1997){M\"urset}, {Wolff}, \&
  {Jordan}}]{1997A&A...319..201M}
{M\"urset}, U., {Wolff}, B., \& {Jordan}, S. 1997, \aap, 319, 201

\bibitem[{{Nedialkov} {et~al.}(2002){Nedialkov}, {Orio}, {Birkle}, {Conselice},
  {Della Valle}, {Greiner}, {Magnier}, \& {Tikhonov}}]{2002A&A...389..439N}
{Nedialkov}, P., {Orio}, M., {Birkle}, K., {et~al.} 2002, \aap, 389, 439

\bibitem[{{Neill} \& {Shara}(2004)}]{2004AJ....127..816N}
{Neill}, J.~D. \& {Shara}, M.~M. 2004, \aj, 127, 816

\bibitem[{{Ness} {et~al.}(2003){Ness}, {Starrfield}, {Burwitz}, {Wichmann},
  {Hauschildt}, {Drake}, {Wagner}, {Bond}, {Krautter}, {Orio}, {Hernanz},
  {Gehrz}, {Woodward}, {Butt}, {Mukai}, {Balman}, \&
  {Truran}}]{2003ApJ...594L.127N}
{Ness}, J.-U., {Starrfield}, S., {Burwitz}, V., {et~al.} 2003, \apjl, 594, L127

\bibitem[{{\"Ogelman} {et~al.}(1984){\"Ogelman}, {Beuermann}, \&
  {Krautter}}]{1984ApJ...287L..31O}
{\"Ogelman}, H., {Beuermann}, K., \& {Krautter}, J. 1984, \apjl, 287, L31

\bibitem[{{\"Ogelman} {et~al.}(1993){\"Ogelman}, {Orio}, {Krautter}, \&
  {Starrfield}}]{1993Natur.361..331O}
{\"Ogelman}, H., {Orio}, M., {Krautter}, J., \& {Starrfield}, S. 1993, \nat,
  361, 331

\bibitem[{{Orio}(2004)}]{2004RMxAC..20..182O}
{Orio}, M. 2004, in Revista Mexicana de Astronomia y Astrofisica Conference
  Series, p.182

\bibitem[{{Orio} {et~al.}(2001){Orio}, {Covington}, \& {{\"
  O}gelman}}]{2001A&A...373..542O}
{Orio}, M., {Covington}, J., \& {{\" O}gelman}, H. 2001, \aap, 373, 542

\bibitem[{{Orio} \& {Greiner}(1999)}]{1999A&A...344L..13O}
{Orio}, M. \& {Greiner}, J. 1999, \aap, 344, L13

\bibitem[{{Orio} {et~al.}(2003){Orio}, {Hartmann}, {Still}, \&
  {Greiner}}]{2003ApJ...594..435O}
{Orio}, M., {Hartmann}, W., {Still}, M., \& {Greiner}, J. 2003, \apj, 594, 435

\bibitem[{{Orio} {et~al.}(2002){Orio}, {Parmar}, {Greiner}, {{\" O}gelman},
  {Starrfield}, \& {Trussoni}}]{2002MNRAS.333L..11O}
{Orio}, M., {Parmar}, A.~N., {Greiner}, J., {et~al.} 2002, \mnras, 333, L11

\bibitem[{{Orio} \& {Tepedelenlioglu}(2004)}]{2004IAUC.8435....2O}
{Orio}, M. \& {Tepedelenlioglu}, E. 2004, \iaucirc, 8435

\bibitem[{{Osborne} {et~al.}(2001){Osborne}, {Borozdin}, {Trudolyubov},
  {Priedhorsky}, {Soria}, {Shirey}, {Hayter}, {La Palombara}, {Mason},
  {Molendi}, {Paerels}, {Pietsch}, {Read}, {Tiengo}, {Watson}, \&
  {West}}]{2001A&A...378..800O}
{Osborne}, J.~P., {Borozdin}, K.~N., {Trudolyubov}, S.~P., {et~al.} 2001, \aap,
  378, 800

\bibitem[{{Papenkova} {et~al.}(2000){Papenkova}, {Aazami}, \&
  {Li}}]{2000IAUC.7477....2P}
{Papenkova}, M., {Aazami}, A.~B., \& {Li}, W.~D. 2000, \iaucirc, 7477

\bibitem[{{Pietsch} {et~al.}(2005){Pietsch}, {Freyberg}, \& {Haberl}}]{PFH2005}
{Pietsch}, W., {Freyberg}, M., \& {Haberl}, F. 2005, \aap, 434, 483

\bibitem[{{Pietsch} {et~al.}(2004){Pietsch}, {Misanovic}, {Haberl},
  {Hatzidimitriou}, {Ehle}, \& {Trinchieri}}]{2004A&A...426...11P}
{Pietsch}, W., {Misanovic}, Z., {Haberl}, F., {et~al.} 2004, \aap, 426, 11

\bibitem[{{Primini} {et~al.}(1993){Primini}, {Forman}, \&
  {Jones}}]{1993ApJ...410..615P}
{Primini}, F.~A., {Forman}, W., \& {Jones}, C. 1993, \apj, 410, 615

\bibitem[{{Rector} {et~al.}(1999){Rector}, {Jacoby}, {Corbett}, {Denham}, \&
  {RBSE Nova Search Team}}]{1999AAS...195.3608R}
{Rector}, T.~A., {Jacoby}, G.~H., {Corbett}, D.~L., {Denham}, M., \& {RBSE Nova
  Search Team}. 1999, Bulletin of the American Astronomical Society, 31, 1420

\bibitem[{{Riffeser} {et~al.}(2003){Riffeser}, {Fliri}, {Bender}, {Seitz}, \&
  {G{\" o}ssl}}]{2003ApJ...599L..17R}
{Riffeser}, A., {Fliri}, J., {Bender}, R., {Seitz}, S., \& {G{\" o}ssl}, C.~A.
  2003, \apjl, 599, L17

\bibitem[{{Riffeser} {et~al.}(2001){Riffeser}, {Fliri}, {G{\" o}ssl}, {Bender},
  {Hopp}, {B{\" a}rnbantner}, {Ries}, {Barwig}, {Seitz}, \&
  {Mitsch}}]{2001A&A...379..362R}
{Riffeser}, A., {Fliri}, J., {G{\" o}ssl}, C.~A., {et~al.} 2001, \aap, 379, 362

\bibitem[{{Sala} \& {Hernanz}(2005{\natexlab{a}})}]{SH2005a}
{Sala}, G. \& {Hernanz}, M. 2005{\natexlab{a}}, \aap, in press
  (astro-ph/0504353)

\bibitem[{{Sala} \& {Hernanz}(2005{\natexlab{b}})}]{SH2005b}
---. 2005{\natexlab{b}}, \aap, in press (astro-ph/0502092)

\bibitem[{{Samus} {et~al.}(2004){Samus}, {Durlevich}, \& {et
  al.}}]{2004yCat.2250....0S}
{Samus}, N.~N., {Durlevich}, O.~V., \& {et al.} 2004, VizieR Online Data
  Catalog, 2250

\bibitem[{{Shafter} \& {Irby}(2001)}]{2001ApJ...563..749S}
{Shafter}, A.~W. \& {Irby}, B.~K. 2001, \apj, 563, 749

\bibitem[{{Sharov}(1993)}]{1993AstL...19..230S}
{Sharov}, A.~S. 1993, Astronomy Letters, 19, 230

\bibitem[{{Sharov}(1994)}]{1994AstL...20...18S}
---. 1994, Astronomy Letters, 20, 18

\bibitem[{{Sharov} \& {Alksnis}(1991)}]{1991Ap&SS.180..273S}
{Sharov}, A.~S. \& {Alksnis}, A. 1991, \apss, 180, 273

\bibitem[{{Sharov} \& {Alksnis}(1992{\natexlab{a}})}]{1992Ap&SS.188..143S}
---. 1992{\natexlab{a}}, \apss, 188, 143

\bibitem[{{Sharov} \& {Alksnis}(1992{\natexlab{b}})}]{1992Ap&SS.190..119S}
---. 1992{\natexlab{b}}, \apss, 190, 119

\bibitem[{{Sharov} \& {Alksnis}(1994)}]{1994AstL...20..711S}
---. 1994, Astronomy Letters, 20, 711

\bibitem[{{Sharov} \& {Alksnis}(1995)}]{1995AstL...21..579S}
---. 1995, Astronomy Letters, 21, 579

\bibitem[{{Sharov} \& {Alksnis}(1996)}]{1996AstL...22..680S}
---. 1996, Astronomy Letters, 22, 680

\bibitem[{{Sharov} \& {Alksnis}(1997)}]{1997AstL...23..540S}
---. 1997, Astronomy Letters, 23, 540

\bibitem[{{Sharov} \& {Alksnis}(1998)}]{1998AstL...24..641S}
---. 1998, Astronomy Letters, 24, 641

\bibitem[{{Sharov} {et~al.}(1998){Sharov}, {Alksnis}, {Nedialkov}, {Shokin},
  {Kurtev}, \& {Ivanov}}]{1998AstL...24..445S}
{Sharov}, A.~S., {Alksnis}, A., {Nedialkov}, P.~L., {et~al.} 1998, Astronomy
  Letters, 24, 445

\bibitem[{{Sharov} {et~al.}(2000){Sharov}, {Alksnis}, {Zharova}, \&
  {Shokin}}]{2000AstL...26..433S}
{Sharov}, A.~S., {Alksnis}, A., {Zharova}, A.~V., \& {Shokin}, Y.~A. 2000,
  Astronomy Letters, 26, 433

\bibitem[{{Shore} {et~al.}(1996){Shore}, {Starrfield}, \&
  {Sonneborn}}]{1996ApJ...463L..21S}
{Shore}, S.~N., {Starrfield}, S., \& {Sonneborn}, G. 1996, \apjl, 463, L21

\bibitem[{{Stanek} \& {Garnavich}(1998)}]{1998ApJ...503L.131S}
{Stanek}, K.~Z. \& {Garnavich}, P.~M. 1998, \apjl, 503, L131

\bibitem[{{Stark} {et~al.}(1992){Stark}, {Gammie}, {Wilson}, {Bally}, {Linke},
  {Heiles}, \& {Hurwitz}}]{1992ApJS...79...77S}
{Stark}, A.~A., {Gammie}, C.~F., {Wilson}, R.~W., {et~al.} 1992, \apjs, 79, 77

\bibitem[{{Starrfield}(1989)}]{sta89}
{Starrfield}, S. 1989, in Classical Novae (Wiley, New York), p. 39

\bibitem[{{Str{\" u}der} {et~al.}(2001){Str{\" u}der}, {Briel}, {Dennerl},
  {Hartmann}, {Kendziorra}, {Meidinger}, {Pfeffermann}, {Reppin}, {Aschenbach},
  {Bornemann}, {Br{\" a}uninger}, {Burkert}, {Elender}, {Freyberg}, {Haberl},
  {Hartner}, {Heuschmann}, {Hippmann}, {Kastelic}, {Kemmer}, {Kettenring},
  {Kink}, {Krause}, {M{\" u}ller}, {Oppitz}, {Pietsch}, {Popp}, {Predehl},
  {Read}, {Stephan}, {St{\" o}tter}, {Tr{\" u}mper}, {Holl}, {Kemmer},
  {Soltau}, {St{\" o}tter}, {Weber}, {Weichert}, {von Zanthier},
  {Carathanassis}, {Lutz}, {Richter}, {Solc}, {B{\" o}ttcher}, {Kuster},
  {Staubert}, {Abbey}, {Holland}, {Turner}, {Balasini}, {Bignami}, {La
  Palombara}, {Villa}, {Buttler}, {Gianini}, {Lain{\' e}}, {Lumb}, \&
  {Dhez}}]{2001A&A...365L..18S}
{Str{\" u}der}, L., {Briel}, U., {Dennerl}, K., {et~al.} 2001, \aap, 365, L18

\bibitem[{{Supper} {et~al.}(2001){Supper}, {Hasinger}, {Lewin}, {Magnier}, {van
  Paradijs}, {Pietsch}, {Read}, \& {Tr{\" u}mper}}]{2001A&A...373...63S}
{Supper}, R., {Hasinger}, G., {Lewin}, W.~H.~G., {et~al.} 2001, \aap, 373, 63

\bibitem[{{Supper} {et~al.}(1997){Supper}, {Hasinger}, {Pietsch}, {Truemper},
  {Jain}, {Magnier}, {Lewin}, \& {van Paradijs}}]{1997A&A...317..328S}
{Supper}, R., {Hasinger}, G., {Pietsch}, W., {et~al.} 1997, \aap, 317, 328

\bibitem[{{Tomaney} \& {Crotts}(1996)}]{1996AJ....112.2872T}
{Tomaney}, A.~B. \& {Crotts}, A.~P.~S. 1996, \aj, 112, 2872

\bibitem[{{Tuchman} \& {Truran}(1998)}]{1998ApJ...503..381T}
{Tuchman}, Y. \& {Truran}, J.~W. 1998, \apj, 503, 381

\bibitem[{{Turner} {et~al.}(2001){Turner}, {Abbey}, {Arnaud}, {Balasini},
  {Barbera}, {Belsole}, {Bennie}, {Bernard}, {Bignami}, {Boer}, {Briel},
  {Butler}, {Cara}, {Chabaud}, {Cole}, {Collura}, {Conte}, {Cros}, {Denby},
  {Dhez}, {Di Coco}, {Dowson}, {Ferrando}, {Ghizzardi}, {Gianotti}, {Goodall},
  {Gretton}, {Griffiths}, {Hainaut}, {Hochedez}, {Holland}, {Jourdain},
  {Kendziorra}, {Lagostina}, {Laine}, {La Palombara}, {Lortholary}, {Lumb},
  {Marty}, {Molendi}, {Pigot}, {Poindron}, {Pounds}, {Reeves}, {Reppin},
  {Rothenflug}, {Salvetat}, {Sauvageot}, {Schmitt}, {Sembay}, {Short},
  {Spragg}, {Stephen}, {Str{\"u}der}, {Tiengo}, {Trifoglio}, {Tr{\"u}mper},
  {Vercellone}, {Vigroux}, {Villa}, {Ward}, {Whitehead}, \&
  {Zonca}}]{2001A&A...365L..27T}
{Turner}, M. J.~L., {Abbey}, A., {Arnaud}, M., {et~al.} 2001, \aap, 365, L27

\bibitem[{{van den Bergh}(1991)}]{1991PASP..103..609V}
{van den Bergh}, S. 1991, \pasp, 103, 609

\bibitem[{{van den Heuvel} {et~al.}(1992){van den Heuvel}, {Bhattacharya},
  {Nomoto}, \& {Rappaport}}]{1992A&A...262...97V}
{van den Heuvel}, E.~P.~J., {Bhattacharya}, D., {Nomoto}, K., \& {Rappaport},
  S.~A. 1992, \aap, 262, 97

\bibitem[{{Vanlandingham} {et~al.}(2001){Vanlandingham}, {Schwarz}, {Shore}, \&
  {Starrfield}}]{2001AJ....121.1126V}
{Vanlandingham}, K.~M., {Schwarz}, G.~J., {Shore}, S.~N., \& {Starrfield}, S.
  2001, \aj, 121, 1126

\bibitem[{{Williams} {et~al.}(2004){Williams}, {Kong}, {Primini}, {King}, {Di
  Stefano}, \& {Murray}}]{williams2004b}
{Williams}, B.~F., M.~R., {Kong}, A.~K.~H., {Primini}, F.~A., {et~al.} 2004,
  \apj, 609, 735

\bibitem[{{Williams} \& {Shafter}(2004)}]{2004ApJ...612..867W}
{Williams}, S.~J. \& {Shafter}, A.~W. 2004, \apj, 612, 867

\end{thebibliography}

\begin{table*}
\caption[]{\xmm, \chandra\ and ROSAT measurements of \m31\ and \me33\ optical novae. 
	   }
\scriptsize
\begin{tabular}{lrrlrrrrrl}
\hline\noalign{\smallskip}
\hline\noalign{\smallskip}
\multicolumn{3}{l}{\normalsize{Optical nova}} & \multicolumn{4}{l}{\normalsize{X-ray measurements}} \\
\noalign{\smallskip}\hline\noalign{\smallskip}
\multicolumn{1}{l}{Name} & \multicolumn{1}{c}{RA~~~(h:m:s)}  
& \multicolumn{1}{c}{JD$^a$} & \multicolumn{1}{l}{Source name$^b$}
& \multicolumn{1}{c}{$D$} 
& \multicolumn{1}{c}{Observation$^c$} & \multicolumn{1}{c}{JD}
& \multicolumn{1}{c}{Day} & \multicolumn{1}{c}{$L^d$}
& \multicolumn{1}{l}{Comment} \\
& \multicolumn{1}{c}{DEC~(d:m:s)} 
& \multicolumn{1}{l}{2440000+} & &(\arcsec)& & \multicolumn{1}{l}{2440000+} 
& \multicolumn{1}{c}{ID$^c$} & \multicolumn{1}{c}{(10$^{36}$ erg s$^{-1}$)} & \\ 
\noalign{\smallskip}\hline\noalign{\smallskip}
[SI2001] &0:41:53.8 & 8977.5&             &   &c1 (EPIC)&11720.5&2743&$<1.1$&\\
1992-01  &41:07:22  &	    &		  &   &c2 (EPIC)&11906.5&2929&$<1.5$&\\
         &          &	    &		  &   &c3 (EPIC)&12089.5&3112&$<2.2$&\\
	 &          &	    &		  &   &1912 (HRC~I)&12213.5&3236&$<6.7$&\\
         &          &	    &[PFH2005]~191&6.4&c4 (EPIC)&12280.5&3303&$37.0\pm1.7$&SSS\\
	 &          &       &[PFH2005]~191&6.4&s1 (EPIC)&12286.5&3309&$13.1\pm0.7$&SSS\\
\noalign{\medskip}
WeCAPP-  &0:42:33.89&12284.3$^<$&         &    &  c4 (EPIC)&12280.5&-4&$<0.5$&\\
N2002-01 &41:18:23.9&       &             &   &2905 (HRC~I)&12290.7&6&$<9.1$&\\
         &          &	    &		  &$\sim$0.2&2906 (HRC~I)&12427.9&144&$27.5^{+12.0}_{-9.5}$& \\
\noalign{\medskip}
WeCAPP-  &0:42:34.61&12150.9&             &   &c3 (EPIC)&12089.5&-61&$<0.5$&close to\\
N2001-12 &41:18:13.0&	    &		  &   &1575 (ACIS~S)&12187.5& 37&$<1.9$&[PFH2005]~287\\
         &          &	    &   	  &$\sim$1&1912 (HRC~I)&12213.5& 63&$0.6\pm0.3$&\\
         &          &	    &             &2.5&c4 (EPIC)&12280.5&130&$0.7\pm0.1$&SSS\\
\noalign{\medskip}
[CFN87]~2&0:42:35.0 & 5225.0&[PFJ93]~33   &2.3&150006h (HRI) &8099.5& 2874&$8.7\pm2.4$&\\
         &41:13:22  &	    &             &   &600474h (HRI) &9541.5& 4317&$<19.8$&\\ 
         &          &	    &             &   &600475h (HRI) &9555.5& 4331&$<13.6$&\\
         &          &	    &             &   &400780h (HRI) &10086.5& 4862&$<8.6$&\\
\noalign{\medskip}
[SI2001] &0:42:35.5 & 8161.5&             &   &150006h (HRI) &8099.5&  -62&$<8.9$&\\
1990-08  &41:13:48  &	    &[SHP97]~181  &3.0&600068p (PSPC)&8465.5&  304&$63.5\pm3.7$&\\
         &          &	    &             &   &600474h (HRI) &9541.5& 1380&$<11.7$&\\
         &          &	    &             &   &600475h (HRI) &9555.5& 1394&$<8.1$&\\
\noalign{\medskip}
[SI2001] &0:42:40.14&10617.5&             &$\sim$0.6& 268 (HRC~I)&11535.6&918&$10.6\pm3.2$& \\
1997-06  &41:15:46.7&	    &             &   &1575 (ACIS~S)&12187.5&1570&$<3.7$& \\
\noalign{\medskip}
AGPV~1576&0:42:42.08& 9622.5&             &   &c1 (EPIC)&11720.5&2098&$<0.6$&\\
         &41:12:18.0 &	    &		  &   &c2 (EPIC)&11906.5&2284&$<0.6$&\\
         &          &	    &		  &   &c3 (EPIC)&12089.5&2467&$<0.6$&\\
         &          &	    &J004242.1+411218&0.7&1912 (HRC~I)&12213.5&2591&$1.2\pm0.4$&K2002\\
         &          &	    &[PFH2005]~313&3.1&c4 (EPIC)&12280.5&2658&$1.6\pm0.2$&SSS\\ 
\noalign{\medskip}
RJC99    &0:42:43.10& 9963.5&		  &   & 268 (HRC~I)&11535.6&1572&$<7.0$\\
Sep-95   &41:16:04.1&       &		  &   & 309 (ACIS~S)&11697.2&1734&$10.2\pm2.7$&\\
	 &          &	    &		  &   & 310 (ACIS~S)&11728.5&1765&$16.1\pm3.6$&\\
	 &          &	    &		  &   &1854 (ACIS~S)&11923.0&1960&$11.5\pm3.1$&\\
         &          &	    &r1-35        &1.1&1575 (ACIS~S)&12187.5&2224&$13.4\pm1.2$&SSS-HR, DKG2004\\
	 &          &	    &J004243.1+411604&0.2&1912 (HRC~I)&12213.5&2250&$8.8\pm0.8$&K2002\\
	 &          &	    &		  &   &2905 (HRC~I)&12290.7&2327&$22.5^{+12.0}_{-8.5}$&\\
	 \noalign{\medskip}
WeCAPP-  &0:42:43.97&11753.0$^*$&         &   &c1 (EPIC)&11720.5&-32&$<0.7$&\\
N2000-03 &41:17:55.5&	    &		  &   &c2 (EPIC)&11906.5&154&$<1.4$&\\
	 &          &	    &		  &  &1854 (ACIS~S)&11923.0&170&$7.1\pm2.4$&\\
         &          &	    &[PFH2005]~320&1.3&c3 (EPIC)&12089.5&337&$9.5\pm0.7$&SSS\\
         &          &	    &r2-60	  &0.8&1575 (ACIS~S)&12187.5&435&$13.4\pm1.2$&SSS-HR, DKG2004\\
         &          &	    &J004243.9+411755&0.3&1912 (HRC~I)&12213.5&461&$8.4\pm0.8$&K2002\\
         &          &	    &[PFH2005]~320&1.3&c4 (EPIC)&12280.5&528&$5.5\pm0.4$&SSS\\
\noalign{\medskip}
WeCAPP-  &0:42:47.45&11719.7$^<$&            &   & 310 (ACIS~S)&11728.5&  9&$<2.2$&\\
N2000-05 &41:15:07.6&       &                &   &1854 (ACIS~S)&11923.0&203&$<4.2$&\\
	 &          &	    &		     &   &1570 (HRC~I)&12071.4&352&$16.8^{+9.7}_{-7.1}$&\\
	 &          &	    &      r2-61     &1.7& 1575 (ACIS~S)&12187.5&468&$9.6\pm1.0$&SSS-HR, DKG2004\\
         &          &	    &J004247.4+411507&0.2&1912 (HRC~I)&12213.5&494&$5.2\pm0.6$&K2002\\
\noalign{\medskip}
[SI2001] &0:42:50.5 &10691.5&             &   &c1 (EPIC)&11720.5&1029&$<1.7$&\\
1997-09  &41:07:48  &	    &		  &   &c2 (EPIC)&11906.5&1215&$<1.6$&\\
         &          &	    &		  &   &c3 (EPIC)&12089.5&1398&$<1.2$&\\
	 &          &	    &		  &   &1912 (HRC~I)&12213.5&1522&$<3.7$&\\
         &          &	    &[PFH2005]~347&2.3&c4 (EPIC)&12280.5&1589&$0.7\pm0.2$&SSS\\
\noalign{\medskip}
GCVS-M31-&0:42:55.2 &10307.5&             &   &c1 (EPIC)&11720.5&1413&$<0.5$&recurrent nova \\
V0962    & 41:20:46 &	    &		  &   &c2 (EPIC)&11906.5&1599&$<1.1$&\\
         &          &	    &		  &   &c3 (EPIC)&12089.5&1782&$<1.0$&\\
	 &          &	    &		  &   &1575 (ACIS~S)&12187.5&1880&$2.2\pm0.8$&\\
         &          &	 &J004255.3+412045&1.2&1912 (HRC~I)&12213.5&1906&$2.9\pm0.8$&K2002\\
         &          &	    &[PFH2005]~359&1.1&c4 (EPIC)&12280.5&1973&$3.4\pm0.2$&SSS\\
\noalign{\medskip}
[SI2001] &0:42:59.3 &10049.5&		  &   &400780h (HRI)&10086.5&  37&$<5.8$&\\
1995-05  &41:16:42  &	    &		  &   & 268 (HRC~I)&11535.6&1486&$7.7\pm2.9$&\\
	 &          &	    &		  &   & 309 (ACIS~S)&11697.2&1648&$8.9\pm2.6$&\\
	 &          &	    &[PFH2005]~369&2.1&c1 (EPIC)&11720.5&1671&$2.3\pm0.5$&SSS\\
	 &          &	    &		  &   & 310 (ACIS~S)&11728.5&1679&$13.8\pm3.3$&\\
	 &          &	    &[PFH2005]~369&2.1&c2 (EPIC)&11906.5&1857&$3.0\pm0.8$&SSS\\
	 &          &	    &		  &   &1854 (ACIS~S)&11923.0&1874&$7.1\pm2.4$&\\
         &          &	    &[PFH2005]~369&2.1&c3 (EPIC)&12089.5&2040&$4.7\pm0.5$&SSS\\
         &          &	    &r2-63        &0.8&1575 (ACIS~S)&12187.5&2138&$11.4\pm1.0$&SSS-HR, DKG2004\\
         &          &	 &J004259.3+411643&1.3&1912 (HRC~I)&12213.5&2164&$10.5\pm0.9$&K2002\\
         &          &	    &[PFH2005]~369&2.1&c4 (EPIC)&12280.5&2231&$7.6\pm0.4$&SSS\\
	 &          &	    &		  &   &2906 (HRC~I)&12427.9&2378&$16.8^{+9.7}_{-7.1}$&\\	 
\noalign{\smallskip}
\hline
\noalign{\smallskip}
\end{tabular}

\label{novae}
\normalsize
\end{table*}

\begin{table*}
\addtocounter{table}{-1}
\caption[]{continued. 
	   }
\scriptsize
\begin{tabular}{lrrlrrrrrl}
\hline\noalign{\smallskip}
\hline\noalign{\smallskip}
\multicolumn{3}{l}{\normalsize{Optical nova}} & \multicolumn{4}{l}{\normalsize{X-ray measurements}} \\
\noalign{\smallskip}\hline\noalign{\smallskip}
\multicolumn{1}{l}{Name} & \multicolumn{1}{c}{RA~~~(h:m:s)}  
& \multicolumn{1}{c}{JD$^a$} & \multicolumn{1}{l}{Source name$^b$}
& \multicolumn{1}{c}{$D$} 
& \multicolumn{1}{c}{Observation$^c$} & \multicolumn{1}{c}{JD}
& \multicolumn{1}{c}{Day} & \multicolumn{1}{c}{$L^d$}
& \multicolumn{1}{l}{Comment} \\
& \multicolumn{1}{c}{DEC~(d:m:s)} 
& \multicolumn{1}{l}{2440000+} & &(\arcsec)& & \multicolumn{1}{l}{2440000+} 
& \multicolumn{1}{c}{ID$^c$} & \multicolumn{1}{c}{(10$^{36}$ erg s$^{-1}$)} & \\ 
\noalign{\smallskip}\hline\noalign{\smallskip}
[SI2001] &0:43:00.7 & 8977.5&             &   &150006h (HRI) &8099.5&-878&$<9.4$&\\
1992-04  &41:14:39  &       &             &4.1&600474h (HRI)& 9541.5& 564&$13.1\pm4.0$&\\
	 &          &	    &		  &2.9&600475h (HRI)& 9555.5& 578&$11.8\pm3.3$&\\
	 &          &	    &		  &   &600674h$^*$ (HRI)& 9926.5& 949&$<40.0$&\\
	 &          &	    &		  &   &400780h (HRI)&10086.5&1109&$<6.3$&\\
\noalign{\medskip}
RJC99	 &0:43:06.96&11019.5&             &   &c1 (EPIC)&11720.5&701&$<2.6$&symbiotic?\\
Jul-98   & 41:18:09.9&      &             &   &c2 (EPIC)&11906.5&887&$<4.3$&\\
         &          &	    &[PFH2005]~395&1.9&c3 (EPIC)&12089.5&1070&$0.9\pm0.4$&SSS\\
         &          &	    &r3-115       &1.1&1575 (ACIS~S)&12187.5&1168&$5.0\pm0.7$&SSS-HR, DKG2004\\
         &          &    &J004306.9+411810&0.6&1912 (HRC~I)&12213.5&1194&$3.8\pm0.9$&K2002\\
         &          &	    &[PFH2005]~395&1.9&c4 (EPIC)&12280.5&1261&$20.1\pm0.7$&hard spectrum, see Fig.~\ref{spec_395}\\
	 &          &	    &		  &   &2905 (HRC~I)&12290.7&1271&$80.0\pm15.0$&\\
	 &          &	    &		  &   &2906 (HRC~I)&12427.9&1418&$<8.4$&\\ 
\noalign{\medskip}
[SI2001] &0:43:16.2 & 8235.5&             &   &150006h (HRI) &8099.5& -136&$<20.1$&\\
1990-12  &41:23:38  &	    &		  &   &600067p (PSPC)&8464.5&229&$<37.1$& \\
         &          &	    &[SHL2001]~230&7.4&SII-E1 (PSPC) &8840.5& 605&$38.9\pm7.7$& strong soft component\\
         &          &	    &[SHL2001]~230&7.4&SII-E2 (PSPC) &8990.5& 755&$42.7\pm5.3$& strong soft component\\
	 &          &	    &[SHL2001]~230&7.4&SII-E3 (PSPC) &9190.5& 955&$17.8\pm5.3$& strong soft component\\
	 &          &	    &		  &   &600474h (HRI)& 9541.5&1306&$<20.9$&\\
	 &          &	    &		  &   &600475h (HRI)& 9555.5&1320&$<29.2$&\\
	 &          &	    &		  &   &400780h (HRI)&10086.5&1851&$<9.7$&\\
\noalign{\medskip}
WeCAPP-  &0:43:18.61&12094.6$^<$&         &   &c3 (EPIC)&12089.5&  -5&$<0.9$&\\
N2001-08 &41:09:49.1&       &J004318.5+410950&1.2&1912 (HRC~I)&12213.5&119&$5.3\pm2.0$&K2002\\
         &          &	    &		  &   &c4 (EPIC)&12280.5& 186&$<0.5$&\\
\noalign{\medskip}
GCVS-M31-&0:43:28.76&10970.5&             &   &c1 (EPIC)&11720.5&750&$<1.0$&recurrent nova \\
V1067    &41:21:42.6&       &             &   &c2 (EPIC)&11906.5&936&$<1.7$& \\
         &          &	    &[PFH2005]~456&1.1&c3 (EPIC)&12089.5&1119&$1.3\pm0.4$&SSS \\
	 &          &	    &		  &   &1912 (HRC~I)&12213.5&1243&$<5.0$&\\
         &          &	    &[PFH2005]~456&1.1&c4 (EPIC)&12280.5&1310&$1.7\pm0.3$&SSS \\
\noalign{\medskip}
[SI2001] &0:43:33.5 & 8280.5&		  &    &600068p (PSPC)&8465.5&185&$<14.8$& \\
1991-12  &41:16:22  &	    &[SHL2001]~246&10.5&SII-E1 (PSPC) &8840.5& 560&$88.3\pm10.1$& strong soft component\\
        &          &	    &[SHL2001]~246&10.5&SII-E2 (PSPC) &8990.5& 710&$51.4\pm6.2$& strong soft component\\
	 &          &	    &		  &    &SII-E3 (PSPC) &9190.5& 910&$<5.8$& \\
\noalign{\medskip}
[SI2001] &0:44:04.71& 8151.9&[SHP97]~268  &   &150006h (HRI)  & 8099.5& -52&$<28.7$&\\
1990-16  &41:18:21.7&	    &		  &2.0&600067p (PSPC) & 8464.5&  313&$12.1\pm1.9$& SSS, see Nedialkov \\
         &          &	    &		  &   &SII-E1 (PSPC)  & 8840.5&  689&$<8.2$&~~~et al. (2002)\\
         &          &	    &		  &   &SII-E2 (PSPC)  & 8990.5&  839&$<2.9$& \\
\noalign{\medskip}
ShAl 57  &0:45:50.17&10718.5&[SHP97]~319  &7.1&600066p (PSPC) & 8463.5&-2255&$9.6\pm2.9$& strong soft component\\
         &41:36:06.1&	    &		  &   &SII-E1 (PSPC)  & 8840.5&-1878&$<7.5$&\\
         &          &	    &		  &   &SII-E2 (PSPC)  & 8990.5&-1728&$<1.7$& \\
         &          &	    &		  &   &SII-E3 (PSPC)  & 9190.5&-1528&$<13.6$& \\
\noalign{\bigskip}
[WS2004] &1:33:47.1 & 9955.5&             &   &400460h-1 (HRI)& 9911.5&  -44&$<8.5$&Nova in M~33 \\
1995-1   &30:40:26  &	    &[HP2001]~93  &1.3&600911h   (HRI)&10110.5&  145&$40.3\pm4.3$& \\
         &          &	    &		  &   &600911h-1 (HRI)&10286.5&  331&$<8.5$& \\
         &          &	    &		  &   &400903h   (HRI)&10460.5&  505&$<8.5$& \\
\noalign{\bigskip}
[WS2004] &1:34:06.9 &10012.5&             &   &400460h-1 (HRI)& 9911.5&  -101&$<3.7$&Nova in M~33 \\
1995-3   &30:37:18  &	    &             &   &600911h   (HRI)&10110.5&    98&$<11.1$& \\
         &          &	    &		  &   &600911h-1 (HRI)&10286.5&   274&$<13.0$& \\
         &          &	    &		  &   &400903h   (HRI)&10460.5&   448&$<2.6$& \\
         &          &       &             &   &0102640101 (EPIC)&11760.5& 1748&$1.6\pm0.8$& \\
	 &          &       &             & $\sim$0.8&786 (ACIS S)&11786.5& 1774&$1.8\pm0.4$& SSS\\
         &          &       &             &   &0102640701 (EPIC)&12095.5& 2083&$<4.0$& \\
         &          &       &             &   &0102642101 (EPIC)&12299.5& 2287&$<1.2$& \\
\noalign{\smallskip}
\hline
\noalign{\smallskip}
\end{tabular}

\label{rest_novae}
Notes: \hspace{0.3cm} $^a$: $^*$ indicates well defined date of optical maximum, $^<$ and $^>$ maximum before or after, else badly defined (see text) \\
\hspace*{1.cm} $^b$: full source names from  K2002 are CXOM31~Jhhmmss.s+ddmmss \\
\hspace*{1.cm} $^c$: for \xmm\ c1 corresponds to ObsID 0112570401, c2 to 0112570601, c3 to 0112570701, c4 to 0112570101 and s1 to 
0112570201\\ 
\hspace*{1.3cm}for \chandra\ we give ObsID and camera used (see text)\\
\hspace*{1.3cm}for the second ROSAT survey  SII-E1, SII-E2, SII-E3 indicate the different epochs (see text)\\
\hspace*{1.cm} $^d$: unabsorbed luminosity in 0.2--1.0 keV band assuming a 50~eV 
black body spectrum with Galactic absorption; 
upper limits are 3$\sigma$ \\
\hspace*{1.cm} $^*$: merged with ROSAT HRI observation 600675h
\normalsize
\end{table*}

\end{document}